\documentclass{article}
\usepackage{graphicx} 
\usepackage{tabto}

\usepackage{tikz,lipsum,lmodern}
\usepackage[most]{tcolorbox}

\usepackage[sorting=none]{biblatex}
\bibliography{hyperrefs}

\title{Towards Responsibly Governing AI Proliferation}
\author{Edward Kembery \\\\ Supervisor: Dr John Burden}
\date{July 2024}

\begin{document}

\maketitle

\begin{center}
    This dissertation was originally submitted for the degree of Master of Philosophy at the University of Cambridge, in July of 2024, and has been lightly edited for upload on arXiv. Readers should consult more recent publications for updates on AI developments.
\end{center}

\vspace{4mm}

\begin{abstract}
   This paper argues that existing governance mechanisms for mitigating risks from AI systems are based on the `Big Compute' paradigm---a set of assumptions about the relationship between AI capabilities and infrastructure---that may not hold in the future.
   \vspace{2mm}

   To address this, the paper introduces the `Proliferation' paradigm, which anticipates the rise of smaller, decentralized, open-sourced AI models which are easier to augment, and easier to train without being detected. It posits that these developments are both probable and likely to introduce both benefits and novel risks that are difficult to mitigate through existing governance mechanisms.
   
\vspace{2mm}
   The final section explores governance strategies to address these risks, focusing on access governance, decentralized compute oversight, and information security. Whilst these strategies offer potential solutions, the paper acknowledges their limitations and cautions developers to weigh benefits against developments that could lead to a `vulnerable world'.
\end{abstract}
\newpage
\tableofcontents
\newpage
\section{Introduction}

\begin{quote}
    \textit{All the committees, the politicking and the plans would have come to naught if a few unpredictable nuclear cross sections had been different from what they are by a factor of two.} 
    \\ \tabto{12mm} - Emilio Segrè, on the making of the atomic bomb \cite[pg. 1] {rhodes2012making}
\end{quote}

\noindent New science enables new technologies; these technologies bear new affordances; those affordances create risks; and those risks inform the mechanisms of mitigating them. But as Segrè's comment suggests, not all risks are equally straightforward to mitigate. As recent philosophers have noted \cite{bostrom2019vulnerable}, if nuclear weapons were easier to develop, or more widely available, it would be far harder to mitigate the risks posed effectively in a manner in keeping with current ethical values. In a rapidly changing scientific field, decision-makers that think they are dealing with one type of `bomb' may find themselves quickly dealing with another. Strategies based on the old paradigm may fail to mitigate the risks of the new. 

The central contention of this paper is that contemporary AI technologies, specifically those based on the well-celebrated generative pre-trained transformer (GPT) architecture (henceforth, `AI'), are rapidly diverging from and challenging traditional assumptions about the development of AI. The first contribution of this paper is to set out this forthcoming paradigm shift; the second, to suggest strategies for mitigating the emerging risks both ethically and effectively.

What sort of shift is proposed? The established AI paradigm emphasises the correlation between AI capabilities, the attendant risks, and the immense number of computational operations required to train and run the system, or ‘compute’ \cite{sevilla2022compute, sastry2024computing}. In practise, this means focusing on risks from powerful frontier models which are closed-weight, high profile, and trained and run on large, trackable compute clusters (e.g. \cite{anderljung2023frontier, heim2024training}). Following Giovanni Dosi's theory of technological “paradigms", \cite{dosi1982technological}, this paper calls this the `Big Compute' paradigm. It permits clear mechanisms for mitigating risks from AI systems responsibly \cite{sastry2024computing} which inform governance decisions from company policy (e.g \cite{anthropic2023scaling, googlesafety, openaiprepared}) to international legislation (e.g. \cite{EUAIAct}).

Today, however, scientific developments are diverging from and challenging these assumptions. Five interoperating technology pathways---Small models, Hidden models, Augmented models, Decentralized processes and Open-Weight models (`SHADOW')--increasingly threaten, like Segrè's cross-sections analogy, to put governance mechanisms out by a factor of two. This paper calls this the `Proliferation' paradigm of AI, or AI proliferation for short. It describes a developing network of technologies and capabilities that is less visible to regulators, harder to control, and presents new risks for irreversible harms. By coining the term, and defining the contours of the Proliferation paradigm, this paper hopes to make a contribution to future research towards modelling and mitigating the risks of models beyond the compute-intensive frontier. 

What sort of governance strategies will work in this new paradigm? Instead of focusing just on compute, governance should focus on the other aspects of what has been termed the `AI triad'---algorithms and informational inputs like data and capability keys---as well as securing non-traditional decentralized compute. Building ethical policy here will require serious and critical attention to the ethical trade-offs of regulation and strategies to mitigate them, as well as improved mechanisms for securing these promises. Only a clear view of the technological paradigm, an evolving picture of stakeholder values, and the mechanisms to secure policy promises will be sufficient to govern the proliferation of AI.

This paper builds its contribution across three main sections. The first sets out key concepts: the argument for governing AI risks; how AI operates in paradigms; and how scientific and technological assumptions have configured Big Compute. The second sets out the Proliferation paradigm by assessing the viability, risks and benefits of each of the SHADOW technologies, drawing attention to how they challenge (compute) governance strategies. The final section explores promising strategies for governing algorithms, decentralized compute and capability keys, paying critical attention to ethical trade-offs of governance and future research priorities. A conclusion positions these policy recommendations in a broader context of AI development, philosophy of science and existential risk. 

\newpage

\section{Key Concepts: Risk \& Governance, Paradigms \& `Big Compute'}

\subsection{The Risk-based Case for AI Governance}

\begin{quote}
    \textit{AI governance refers (1) descriptively to the policies, norms, laws, and institutions that shape how AI is built and deployed, and (2) normatively to the aspiration that these promote good decisions (effective, safe, inclusive, legitimate, adaptive).} 
    \\ \tabto{18mm} - Allan Dafoe, `AI Governance: A Research Agenda' \cite{dafoe2024ai}
\end{quote}

In its broadest sense, `AI governance’ simply refers to set of operations designed to affect and improve decisions about how AI models are built and deployed across society \cite{dafoe2024ai}. It might refer to decisions in private companies \cite{schneider2020ai}, national governments \cite{taeihagh2021governance}, or international organisations \cite{ho2023international}. It might also refer to decision frameworks designed to achieve very different goals. Governance frameworks built to maximise the capabilities of AI systems overall, for example, might be very different to those built to minimise the likelihood of an adversarial nation acquiring those capabilities.  The analysis in this paper focuses on two priorities necessary for responsible governance: mitigating harms from AI systems effectively, and doing so ethically. 

The possibility of severe harms from powerful AI is a well-established prospect \cite{anderljung2023frontier, shevlane2023model, brundage2018malicious}. For instance, highly capable models could assist in the creation and deployment of biological weapons \cite{sandbrink2023artificial, openai2024bio} or cyberattacks against organisations or critical infrastructure \cite{mirsky2023threat, malatji2024artificial}. They could also experience technical failure modes like cybersecurity vulnerabilities or misalignment \cite{ji2023ai}, either at the level of individual models, or those in larger groups \cite{anwar2024foundational}. Even if models are used with good intentions, some applications, like automation, may create social harms in which disproportionately disempower some portions of the labour market more than others \cite{acemoglu2021harms}. Even if they were willing, it is unlikely that corporations alone would be able to mitigate such risks \cite{slee2020incompatible}. Effective countermeasures are likely to require coordination from both private and public actors (e.g. \cite{moulange2023responsible}).

At the same time, it is crucial that both AI technologies and governance interventions function \textit{ethically}: that is, that they strive to reflect the normative aspirations of the parties which make and experience the results of the intervention \cite{dafoe2024ai}. That might include implementing governance measures which prioritise certain definitions of fairness \cite{richardson2021framework}, certain varieties of privacy \cite{veliz2020data, Veliz2020-VLIPIP}, clear lines of accountability \cite{novelli2023accountability}, or support certain definitions of democracy \cite{seger2023democratising}, especially inter-generational  \cite{greaves2021case} and international perspectives \cite{roberts2024global, daly2020ai}. Finding a `one size fits all solution' is not the goal here. Rather, the aim should be to encourage constructive debate around the values \textit{necessarily} embedded in AI governance regimes so as to work progressively towards generally held principles.

A pervading concern is where to set the \textit{limits} of such governance. As Zeng (2020) and others have noted, over-regulation could enable and enact the centralization and predominance of some values or actors over others, compromising broadly liberal democratic values \cite{zeng2020artificial, dafoe2018ai}. Although empirical studies might go some way towards illuminating these trade offs, AI systems present significant unknowns, and policy must necessarily make values-driven interpretations bearing inductive risks \cite{douglas2000inductive}. Developing an ethical framework for making these values-based decisions (for instance, involving democratic processes) is consequently also a concern. Great as the risks from AI systems are, over-regulation resulting from unwarranted interpretations based on scarce information might also bear unethical consequences. These issues will be discussed in more detail in section 4.

\subsection{The `Big Compute' Paradigm}

\begin{quote}
    \textit{Computing power, or “compute", is crucial for the development and deployment of artificial intelligence (AI) capabilities. As a result, governments and companies have started to leverage compute as a means to govern AI.} 
        \\ - Sastry et al., `Computing Power and the Governance of Artificial Intelligence' (2024) \cite{sastry2024computing}
\end{quote}

Understanding how AI proliferation challenges established assumptions about how to mitigate the risks from AI technologies requires first understanding what those established assumptions are. A useful lens here is the `technological paradigm', an embodied intellectual construct Giovanni Dosi defines, following Kuhn, as “a `model' and a `pattern' of solution of \textit{selected} technological problems, based on \textit{selected} principles derived from natural sciences and on \textit{selected} material technologies" (\textit{sic}) \cite[p. 150]{dosi1982technological}, including the “industrial structures" (labs, firms, government institutions) associated [p. 157]. For Dosi, such paradigms were described by `technological trajectories', intellectual and industrial progress towards the solution of paradigmatic problems “whose outer boundaries [define and] are defined by the nature of the paradigm itself" (p. 154). Well-established across various fields (e.g. \cite{toufaily2021framework, ciarli2021digital, ho2015typology}), this lens can help decision makers to appreciate technological paradigms like webs of interconnecting assumptions: grounded in evolving scientific research, and embodied not only in experimental and commodity technologies, but the corollary industries, corporations, and community practises of those researching and developing them. 

Dosi's `technological paradigm' provides a useful lens to understand AI today. Of the three facets necessary to develop powerful AI capabilities---informational inputs like data, algorithms, and compute (sometimes referred to as the `AI triad' \cite{buchanan2020triad})---compute has been assigned the dominant position. This is for good reasons. The hypothesis that performing more operations in the training and inference stages of a model’s operation will improve that model’s performance is one of the most well-supported findings in machine learning \cite{hoffmann2022training, sevilla2022compute, owen2024predictable}. This means that models involving a larger number of parameters, or those trained for longer, typically perform better than smaller models \cite{hoffmann2022training}. Companies capable of gathering more compute can therefore build more capable models. This training process is critically expensive: Sastry et al. (2024) point out that “development of frontier AI systems has become increasingly synonymous with large compute budgets, access to large computing clusters, and the proficiency to leverage them effectively” \cite[p.9]{sastry2024computing, Situational}.

This bears a number of corollaries for how the AI industry has developed. Since most leading AI companies rely on colossal personal datacentres or those of high-visibility compute providers or `hyperscalers’ \cite{panjwani2020study}, capable models are difficult to hide.\footnote{For context, a two-order-of-magnitude increase in compute, comparable to the jump between GPT2 and GPT4, would require the same power footprint as the Hoover Dam \cite{Situational}.} Each order-of-magnitude jump in investment here is extremely expensive. Training a model on one order of magnitude more compute than GPT4 might cost billions of dollars \cite{Situational}. Releasing information about such a model’s architecture, weights and training data online for free without usage constraints would be financial suicide (or at the very least a huge advantage to competing companies). Beginning with the compute hypothesis, one arrives at a paradigm of powerful models which are difficult to replicate, high profile, and trained and run on large, expensive, trackable compute clusters. The spiritual successor to `Big Data''s emphasis on scale \cite{martin2020ethical}, this paper refers to this paradigm as `Big Compute'.

The 'Big Compute' paradigm in turn configures what Wirtz, Langer and Weyerer (2024) have called the regulatory "ecosystem" \cite{wirtz2024ecosystem}. It bears three advantageous features for responsible risk-mitigation governance. First, it creates \textit{moats}: huge datacenters are expensive to run and maintain, so there are relatively few actors creating frontier models, lowering the burden on regulators and ensuring against over-regulation of non-threatening actors \cite{egan2023oversight}. Second, it is \textit{anticipatory}: the amount of compute used for training runs serves as a useful proxy for capabilities, allowing policymakers to better assess which models are and are not dangerous and decide when to trigger further evaluations \textit{before} models are released or fully developed. One such governance mechanism, “compute thresholds" \cite{heim2024training}, forms a critical part of both the US AI Executive Order \cite{biden2023executive} and the EU AI Act \cite{EUAIAct}, and some AI companies deploy them as part of “responsible scaling policies" (e.g. \cite{anthropic2023scaling}). Third, compute is \textit{physical}: more specifically, as Sastry et al. argue in their seminal paper on compute governance, it is “detectable, excludable, and quantifiable, and produced via an extremely concentrated supply chain” \cite[p.1]{sastry2024computing}. This helps to ensure transparency, enable reallocation of resources, and “enforce restrictions against irresponsible or malicious AI development and usage” (p.1). 

Taken as a whole, the `Big Compute' paradigm represents a particular set of scientific hypotheses, a set of technological trajectories, a narrow set of key actors, and an advantageous paradigm for governance. It is empirically well-founded, and likely to be an important tool for structuring effective risk-mitigation governance mechanisms from large and frontier models in the future. Nonetheless, as Sastry et al. themselves note, “governance of compute is not the whole story of AI governance" \cite[p. 2]{sastry2024computing}. That AI will remain bound to these assumptions as the technology matures is far from guaranteed. The next section will explore some of the scientific research and technological developments that offer an supplementary paradigm to the contemporary model of `Big Compute', and their associated risks and governance challenges.

\section{The Proliferation Paradigm: `SHADOW' Pathways}

\begin{quote}
    \textit{The success of a paradigm ... is at the start largely a promise of success discoverable in selected and still incomplete examples.} 
    \\ \tabto{5mm} - Thomas Kuhn, \textit{The Structure of Scientific Revolutions} \cite[p.23]{kuhn1997structure}
\end{quote}

The central contention of this paper is that recent scientific and technological developments increasingly interoperate to provide and embody alternative assumptions to those of `Big Compute'. Taken together, one could call this the `Proliferation' paradigm of AI, or AI proliferation for short. It proposes serious new challenges for risk-mitigation which will be explored in Section 4.

A robust technological description of this paradigm has yet to be made. To this end, this paper contributes the `SHADOW' framework, a map of five emerging pathways---Small models, Hidden models, Augmented models, Decentralized processes and Open-Weight models (`SHADOW')---which help define the contours of the Proliferation paradigm. Each proceeding subsection sets out the technical viability of the pathway, assesses its risks and opportunities, and describes how it challenges a governance paradigm based on compute, thus setting out the landscape for risk-mitigation governance that will be explored in Section 4.

\subsection{Small Models} 

The more compute required for models to demonstrate dangerous capabilities, the more effective governance via compute becomes. Yet such a framework might overlook small models: specifically, those which demonstrate dangerous capabilities, yet have far fewer parameters or training token requirements, requiring little compute to train, run and fine-tune. This paper defines ‘small’ in a relative sense:

\begin{itemize}
    \item Fewer than large frontier labs, such that compute governance thresholds designed to catch frontier systems do not apply.
    \item Few enough to make (the cost of) creating, running and fine-tuning them accessible to a dramatically larger pool of people.\newline
\end{itemize} 

\textbf{Technical Pathway(s)} \newline

\noindent The algorithms behind AI systems may become far more efficient over time.\footnote{The efficiency of an algorithm can be understood as the number of floating-point operations required to demonstrate a particular capability, where more efficient algorithms use less compute \cite[p. 9]{scharre2024future}.} As Brown and Hernandez demonstrated in their 2020 paper popularising the idea, “the number of floating-point operations required to train a classifier to AlexNet-level performance on ImageNet decreased by a factor of 44x between 2012 and 2019”, corresponding to a doubling in efficiency “every 16 months over a period of 7 years” \cite{hernandez2020measuring}. State of the art models will most likely shrink faster: Ho et al. conclude that language model efficiency is doubling on average every 8.4 months (albeit with a broad 95 percent confidence interval of 5.3 to 13 months) \cite{ho2024algorithmic}.  Technical strategies like quantization \cite{gholami2021survey}, model pruning \cite{cheng2023survey}, knowledge distillation \cite{kaleem2024comprehensive}, and parameter-efficient fine-tuning \cite{xu2023parameterefficient} have all been cited as promising avenues for diminishing the amount of compute required to train a model, or the parameters needed, for a model to achieve advanced capabilities. 

Two technical pathways to smaller models stand out. One, best demonstrated by Microsoft's Phi series, trained on a data corpus of real and synthetic (GPT 3.5 generated) textbook questions, matches or outperforms models like Gemma 7B and Llama 8B on benchmarks despite possessing about half the number of parameters \cite{abdin2024phi}.\footnote{`3B' refers to 'three billion parameters', a proxy for the compute used to train the model. OpenAI's GPT3, released in 2020, possessed 175 billion parameters \cite{brown2020language}.} Apple’s OpenELM 3B model, uses such a specialised training regime to score a 92.7 on SciQ, an undergraduate level science benchmark \cite{mehta2024openelm}. Secondly, improvements to algorithmic architectures have also seen models become more capability dense \cite{shazeer2017outrageously}. Models like Databricks’ DBRX 132B and Artic, a 40B parameter model with 17B parameters active, use only a fraction of its parameters on any given query, improving inference efficiency \cite{databricks2024dbrx, artic2024snowflake}. Such systems work by allowing models to `search' through individual subnetworks; improving the ability of models to allocate compute strategically might allow developers to make them even smaller \cite{bittererlesson}. Deeper architectural changes, like a shift from MLPs to Kolgomorov-Arnold Networks, might further reduce the compute requirements \cite{liu2024kan}. If frontier AI models become capable of automating AI research and development and improving algorithmic efficiency (as suggested in e.g. \cite{song2024position}), this might create a ‘slipstream effect’, by which growing scale of frontier models paradoxically rapidly \textit{decreases} the compute required to achieve dangerous capabilities.

Skepticism is necessary here. Timelines are unclear: whilst it could be that models get smaller and better quickly (see section 3.3), unknown ceilings may cause progress to plateau. Whether improving frontier models make building small models easier, or less financially attractive, and how this will affect the pathway development, is also currently unclear. Nonetheless, it seems likely that some model shrinking will occur in the short term, increasing the number of actors capable of building models with sub-frontier capabilities, a phenomenon Pilz and Heim refer to as `diffusion' \cite{pilz2023increased}. \newline

\textbf{Benefits, Risks, and Governance Challenges} \newline

\noindent Small models hold real promise. As a recent paper notes, the current size of larger models makes them unsuitable for scenarios requiring “on-device processing, energy efficiency, low memory footprint, and response efficiency” \cite{thawakar2024mobillama}. Smaller models are also suitable for running on-device, protecting them from privacy issues associated with cloud compute providers, and potentially making them easier to audit \cite{mehta2024openelm}. They are also, perhaps most importantly, cheaper to run, suggesting that (as Nathan Lambert argues) they are well positioned to “unlock the most economic value by unlocking applications reliant on edge computing or low-costs” \cite{lambert2024artic}.   

These benefits come alongside new risks. A downside of what Pilz, Heim \& Brown call the ``access affect" \cite{pilz2024increased} is the increased likelihood of malicious actors gaining access to dangerous capabilities. As a Center of New American Security paper points out, within five years, at current trends, “the cost to train a model at any given level of capability decreases roughly by a factor of 1,000, or to around 0.1 percent of the original cost” \cite[p. 30]{scharre2024future}. By this trajectory, training a model in 2029 to today’s state of the art lies in the price range of a family car (based on cost estimates, p. 13).  Assuming similar training costs per parameter, training a model like Microsoft’s Phi 3 the same year would cost around \$83, about the cost of a family restaurant meal. Other technological trajectories discussed in this paper, like the falling price of compute access (section 2.3), capabilities augmentation (2.4) and open-weight ready-mades (2.5), might make them even cheaper.

Dangerously powerful small models present obvious risk-mitigation challenges, especially to a governance paradigm centered on compute. A key loss is the viability of compute thresholds. Sparing intervention by “scientific panel” (51.1.b), models developed in the EU below the threshold need only be deemed safe to release by the company or individuals developing them, a decision for which there is as yet no commonly accepted standard \cite{anwar2024foundational}. Although  proponents of compute governance suggest that these thresholds should adjust over time (e.g. \cite[p. 70]{sastry2024computing}), the evidence and procedures that would be necessary to do so are unclear. Even if model evaluation mechanisms improved substantially to address this threat, another challenge is the increased opportunity for regulatory evasion. Smaller models would also be easier to hide from government oversight based on compute, especially if trained on decentralized architectures (2.3). This is discussed in more detail in the next section. 

\subsection{Hidden Models}

The more information regulators, governments and the public have about which actors are developing and deploying what sort of AI models, the better equipped they are to make responsible decisions about how best to govern AI systems. In contrast, ‘hidden’ models – models with dangerous capabilities which go unreported, unregistered, or unlicensed in the jurisdiction in which they are used, or which are otherwise overlooked by government building AI regulation – present a variety of risks. Whilst not a technological pathway \textit{per se}, hidden models constitute an important and as-yet undocumented aspect of the "industrial structure" of the Proliferation paradigm \cite[p. 157]{dosi1982technological}.\newline

\textbf{Hidden Model Pathway(s)} \newline

\noindent Models might be `hidden' in a variety of ways: call them \textit{below, above}, and \textit{outside}. \textit{Below} would concern the models that are \textit{legitimately} not covered by (`below') compute-threshold based reporting requirements (e.g. any model trained on sub-10\textsuperscript{26} FLOPs in the US) discussed in Section 2.1. This is likely to become more viable as eliciting dangerous capabilities requires less compute (section 2.3 and 2.5). However, it should be noted that the more widely used the model is, the more likely it will come to the attention of regulators. 

\textit{Above} would concern models with dangerous capabilities which are covered by (`above') compute threshold-based regulation, but go inadequately unreported. One culprit here might be major labs. While it is unlikely a leading lab would release an untested model outright, labs can and do release models `into the wild’ covertly for testing purposes. The origins of `gpt2’, briefly the most powerful LLM on the LMSYS leaderboard, were unknown before it was revealed to be GPT4o \cite{techopedia2024gpt2}. If a powerful model was trained and deployed covertly, copied (see `slipstream' effect, section 3.3), or leaked, a massively dangerous model might proliferate in the wild. Even if models do not go entirely unreported, nascent or low-quality reporting might still mean that developers building frontier models fail to prepare \cite{kolt2024responsible}. In the longer term, another might be malicious actors using decentralized compute (see section 3.4) to build powerful models that regulators cannot see.

\textit{Outside} would concern models located outside a jurisdiction, thus `hidden' from the regulation within it, but accessed via mechanisms like VPNs. The present likelihood that this would happen is unclear. On the one hand, countries like China, Saudi Arabia and the UAE have all proposed alternative governance regimes to the EU and US, open-weighting their state-led models like Qi \cite{bai2023qwen}, Allam \cite{IBM2024allam} and Falcon \cite{reuters2023falcon} respectively.  On the other, threatening models may be used within national borders as much as outside them, limiting the likelihood that unstable countries might wish to invest in them. At least two high risk pathways look likely here. One is that a national government uses some form of hidden AI system to perpetrate harms in a warfare context \cite{brundage2018malicious}. The second is that a national government with less robust evaluation procedures mistakenly ratifies the safety of a model that then causes international harm \cite{anderljung2023frontier}. \newline

\textbf{Benefits, Risks, and Governance Challenges} \newline

\noindent Government oversight imposes burdens. Reporting costs time and attention; evaluations are expensive to run and difficult to check. If capabilities can be attained faster without members of small teams having to spend their time reporting to government bureaucrats, with no risk of releasing a model that might cause harms, it stands to reason that they might protest this, especially if their reporting requirements were similar to those of an AI company many times their size \cite[p. 67]{sastry2024computing}. This has been the argument of many ‘pro-innovation’ or ‘anti-regulation’ arguments for governing AI \cite{eiras2024near}. 

At the same time, however, a laissez-faire approach to unreported models could have serious risks (see Section 2). For instance, a model hidden \textit{below} reporting requirements could be exposed to less strict evaluation requirements, leading to the deployment of a model with flaws, like security vulnerabilities, in critical infrastructure. A model hidden \textit{above} requirements governments are taken by surprise by the powerful capabilities of a less-reported model, leading to societal unrest \cite{anwar2024foundational}. And a model hidden \textit{outside} requirements could enable malicious actors to perpetrate serious harms either domestically or in nations with higher-stringency governance regimes \cite{brundage2018malicious}. In each of these cases, having the information necessary to dissolve the threat before manifests would seem to be an urgent priority. 

Governing hidden models presents two main challenges. One is how to calibrate the risk sensitivity when designing evaluation regimes that might affect small actors more than large ones (see section 4). Another is how to protect against regulatory flight. In a `Big Compute' paradigm, AI-capabilities drivers (among other things) are high profile, possess big teams, and require huge data centers. AI proliferation might reduce these requirements, making regulatory flight substantially easier. If governance standards are not global, actors with higher risk thresholds may release models that will undermine the governance regimes established in nations working to establish lower-risk thresholds (see section 4). 

\subsection{Augmented Models}

Where Section 3.1 focused on the diminishing compute requirements of training dangerously capable models, this section focuses on the fleet of strategies for eliciting new capabilities (`capability keys') from AI systems which bypass or invalidate safeguards (`jail-break' models), without retraining. Although a related problem is mentioned by Anderljung et al. (2023) \cite{anderljung2023frontier} as the `unexpected capabilities problem' (Sastry et al. (2024) \cite[p. 34]{sastry2024computing} also note this possibility in a footnote), this section explores more up-to-date literature on these risks in more detail. \newline 

\textbf{Technical Pathway(s)}\newline

\noindent Two technical pathways stand out as promising for augmenting the capabilities of existing models: advanced prompting, and precise fine-tuning. Well-known prompt-based strategies, like chain of thought reasoning \cite{wei2023chainofthought} and prompt scaffolding \cite{suzgun2024metaprompting}, have been shown to have impressive effects on the quality of model outputs. Davidson et al. (2023) report “gains equivalent to training with 5 to 20x more compute at less than 1\% of the cost” \cite{davidson2023ai}. Working in the context window (sometimes referred to in technical circles as `adaptive compute' \cite{dwarkeshtrenton}) looks to become more popular in future. Google's recent Gemini model can learn new languages with efficiency using only context inputs \cite{geminiteam2024gemini}. These techniques, compatible with both open- and closed-source models, might proliferate online, to be used at will. In the style of some video games, one could think of these pieces of information as `capability keys'. 

A complementary possibility is fine-tuning, which has been shown to enable both greater capabilities from small models \cite{han2024parameterefficient} and remove safeguards on existing models with ease \cite{qi2023finetuning}. Model merging, a related technique, allows developers to systematically combine models to create more powerful capabilities, releasing “new state-of-the-art models back to the open-source community” \cite{akiba2024evolutionary}. It is even possible to fine-tune smaller models on the output data of larger ones, leading to a bootstrapping effect which small models become more capable. As in section 3.1, this might create a slipstream effect, by which the improvement in frontier models radically raises the quality of sub-frontier-compute-level models. These model weights could thus also be thought of, in a way, as `capability keys': though on a different scale to prompt-based attacks, and only affecting models with open fine-tuning interfaces like open-weight models (section 3.5). Both strategies---prompt-based and fine-tuning---might also be used together.

A catalysing factor here is the emergence of powerful supporting infrastructure for sharing information about augmenting AI model capabilities. Open-source AI communities HuggingFace, LAION,  or Eleuther AI proudly support the proliferation of state-of-the-art machine learning techniques.  Semi-automated systems, like Cohere’s Command R+ \cite{cohere2024} and Github’s in-platform Copilot \cite{github2021} aim to accelerate coding and bring capabilities within the grasp of less experienced researchers. Microsoft’s AutoGen already designs multiagent systems automatically \cite{wu2023autogen, zhang2024training}, as does Evolutionary Model Merge for model merging \cite{akiba2024evolutionary}. If models were to become substantially better at AI R\&D (as highlighted in DeepMind’s recent frontier safety framework \cite{deepmind2024fsf}), these capabilities might proliferate more widely still. \newline

\textbf{Benefits, Risks, and Governance Challenges}\newline

\noindent Eliciting more powerful capabilities from existing models has obvious benefits. Prompting strategies can make outputs more robust, allowing smaller models to be scaffolded in ways that make them more effective agents \cite{agarwal2024multi}.  Fine-tuning allow models to be cheaply and efficiently adapted for specific use cases, and has been shown to protect against certain failure modes \cite{christiano2023deep}. Infrastructure that helps these techniques proliferate in society more broadly has advantages for decentralizing power over model capabilities \cite{fard2021distributing}; it might also disproportionately empower smaller industry actors \cite{brynjolfsson2023big}, and help to create broader bases for AI talent \cite{eiras2024near}.

However, model augmentation brings hard-to-forsee risks that Anderljung et al. refer to as the `unexpected capabilities problem' \cite[p. 12]{anderljung2023frontier}. If models quickly became extremely powerful due to a slipstream effect, the best case scenario might be societal disruption. Alternatively, if prompt scaffolding compounded capabilities, but created cybersecurity vulnerabilities, it could compromise critical infrastructure. More worryingly, present adversarial prompting \cite{wei2024jailbreak} or fine-tuning \cite{qi2023finetuning} can bypass safeguards with ease, uplifting malicious actor capabilities. A paradigm in which many older unsecured models could be upgraded to possess dangerous capabilities would be an immense and challenging regulatory target, especially if many were open-weighted (section 3.5). How to secure capability keys---which might proliferate online over platforms like HuggingFace or EleutherAI (see Section 3.5)---should be a governance priority. If augmentable models are distributed open-source for actors to download them (section 3.5), or difficult to track (3.2), it would be especially difficult for auditing parties to spot these issues and address them, and for governments to evaluate their spread and impact (see Section 4). 

\subsection{Decentralized Processes}

`Big Compute' generally involves a few high-profile `hyperscalers', working with a few high-profile companies. This section focuses on alternative pathways by which AI models might access the necessary compute, either by accessing decentralized compute providers, or by operating as a decentralized service across many devices in real time. Although decentralization has been widely discussed in a governance context around blockchain, applauded in AI circles \cite{montes2019distributed, kersic2024review}, and briefly featured in a 2020 AI review \cite{gupta2020decentralization}, no up-to-date governance-facing papers on the technology existed the time of writing.\newline 

\textbf{Technical Pathway(s)}\newline

\noindent There are two ways decentralized processes might transform AI. The first involves training a centralised model on a decentralized compute platform; the second, training a decentralized model on lots of individual sites. To understand the first, consider Akash, a “decentralized compute marketplace” or “supercloud” that allows users to buy compute from providers at low costs \cite{akash2023}.\footnote{For a technical review of how decentralized platforms work, see \cite{lin2024decentralized}.} The system already runs small in-context learning models (Llama 7B), and the team have trained at least one large language model (“Thumper”) entirely on the distributed architecture \cite{akash2023}.  Other similar examples include Golem \cite{golem},  Brain Chain \cite{deepbrainchain} and iExec \cite{iexec}.  Other frameworks like Gensyn go further still, implementing an automated payment system over blockchain that allows “direct and immediate rewards for those contributing their computational resources for machine learning tasks” \cite{gensyn2024}. If companies could unlock the `latent compute' already in existence and make it public cheaply, Aksh Garg (a decentralized compute researcher) points out, “even tapping into a tiny fraction of this widely distributed network of computation would be game-changing" \cite{garg2024}.

A second stream might come from distributed models themselves. Douillard et al. (2024) argue that models trained on modular tasks from discrete sites can be more efficient: training a model on the C4 dataset, with paths “of size 150 million parameters”, the team was able to match “the performance in terms of validation perplexity of a 1.3 billion model, but with 45\% less wall clock training time”, which they hope will enable more energy and compute efficient scaling (see Section 2.1) \cite{douillard2024dipaco}. Meanwhile, papers like LinguaLinked promise models that will be able to be run (perform in-context learning) using the compute from several devices simultaneously \cite{zhao2023lingualinked}. As models become smaller (Section 2.1) or more efficient (2.3), this might be an increasingly viable strategy for eliciting greater capabilities from models without relying on traditional cloud compute providers.

At present, decentralized processes do not play a significant role in AI. Decentralized compute networks are very small (Akash offers access to 85 A100 GPUs; a mid-tier AI company like Stability AI might own 5,400).  They are also hamstrung by technical limitation associated with running a model on a virtual machine rather than a data center co-located in the same place \cite{garg2024}. Research into decentralized processes like DiPaCo is still in the theoretical stages. Nonetheless, the amount of investment in the industry is substantial. DeepMind’s investment in distributed path decomposition is illustrative of their confidence that they will be able to create an AI paradigm in which individual “paths, trained on any available hardware type, communicate infrequently across the world, exchanging useful information and enabling new forms of composition” \cite{douillard2024dipaco}. If progress on this frontier continues, perhaps accelerated by the other SHADOW technologies, the risks should be taken seriously. \newline 

\textbf{Benefits, Risks, and Governance Challenges}\newline

\noindent Proponents of decentralized compute networks point to benefits like “lower cost and greater choice”, “access standardisation”, “community driven” governance regimes and income streams for smaller actors \cite{akash2023}. They argue that centralized compute provisions push for longer contracts, enforce unsuitable UX, and drive up prices \cite{gensyn2024}. Distributed models might work better across decentralized compute: they also promise to bring down computing costs, and leverage existing compute more efficiently, helping to make capabilities more widely available \cite{garg2024}.

Two potential risks of a paradigm built around decentralized computing stand out. The first is that this reduces the transparency of model development, leading to sub-optimal policy or harms from unsafe systems (see section 2.2). Yet building pro-transparency policy is challenging, too. Companies like Akash, vocal in their criticism of the “now-largely-discredited… cautious approach to technological progress”, note that “a demand exists for access to open-source models in a permissionless environment” where “anyone can run these models without unnecessary restrictions” \cite{akash2023}. It is unlikely that they would enact pro-transparency Know-Your-Customer policies like traditional cloud computing providers, even if it was technically feasible \cite{egan2023oversight}.  Even then, their privacy-minded client base (both buyers and sellers) might just as easily find another marketplace elsewhere.  

The second, more concerning risk, is that decentralized compute might constitute a resource for malicious actors to access compute to fine-tune or build small models for offensive purposes (see sections 2.1 and 2.3) without raising the suspicion of authorities (see section 2.2). If decentralized compute networks were powerful enough, they might prove a way for malicious actors to get around on-chip governance measures such as location tracking, in effect skirting anti-chip-smuggling policy regimes.  Even if actors are not malicious, they might make it more difficult to enforce policy regimes than traditional physical compute, or large corporations (the decentralized computing provider might not be based in the same jurisdiction as the model is trained, for instance). Distributed systems make enforcement even more difficult: enforcing a model trained on a system like like DiPaCo would compel action against 256 geographically distributed entities in the case of reluctant or malicious actors. Like the capability keys of Section 3.3, once the information necessary to build such systems is publicized, it may be impossible to contain. 

\subsection{Open-Weight Models}

Copying, downloading or using (components of) models that are freely available online uses far less compute than training a model from scratch. The more of these components that are freely available (which might include pre-training data, fine-tuning data, alignment data, evaluation frameworks, model architecture, weights, and their respective implementation code \cite{eiras2024near}), the more a model can be described as ‘open-source’. Since some facets of the training process (eg. the inference code) cost far less compute to generate than others, this paper follows Seger et al. in focusing most on open-weight models “for which at least model architecture and trained weights are publicly available” \cite[p. 2]{seger2023open}.\newline

\textbf{Technical Pathway(s)}

\noindent When Seger et al. (2023) wrote their seminal piece on the risks and benefits of open-sourcing highly capable models, there was yet to be an open-source model with capabilities close to the frontier. Today, open-weight models are a significant and competitive part of the AI ecosystem. They routinely beat frontier models like OpenAI’s GPT4 on the LMSYS arena \cite{chiang2024chatbot}, both in narrow domains (e.g. Cohere’s Command R +, \cite{cohere2024}) and more generally (e.g. Alibaba’s Qwen 1.5 72B \cite{bai2023qwen}). Popular models like Meta’s Llama 3 70B (April 2024) outperform or tie across multiple benchmarks with closed-source models produced by larger companies only a few months before (e.g. Google’s Gemini 1.5, February 2024 or Claude 3 Sonnet, March 2024) \cite{meta2024}. Looking towards Facebook’s Llama 405B, it is possible that around the time of this paper’s publication with see a frontier-leading model be open-weighted for the first time \cite{dwarkesh2024zuck}. 

Open-weighting is an increasingly popular and well-consolidated practise. Huggingface, a platform for downloading and sharing open-source models, supports over 350k models today \cite{huggingfaceindex}.  Other platforms include Open-LAION,  Red Pyjama, and Eleuther AI.  In terms of big players, Meta has committed to open-weighting their frontier models (with some constraints) for the foreseeable future \cite{dwarkesh2024zuck}, and Open-AI has shown a pattern of open-sourcing their models roughly a year after they are first deployed. At the smaller end, Apple’s OpenELM series is one of the most extensively open-sourced models currently available, including training datasets with their model reports \cite{mehta2024openelm}. In lieu of responsible scaling policies with stipulations around open sourcing, there is nothing to stop companies from open-sourcing increasingly large and powerful models.\newline

\textbf{Benefits, Risks, and Governance Challenges}

\noindent Providing the weights of capable models to the public for free has obvious and significant advantages. One concerns the progress of AI systems themselves. As Eiras et al. note in a recent paper (supported by Meta), open-weight probably increases the number of people capable of working in frontier research and contributing to AI development more generally, potentially reduces the need for retraining models at personal expense (thereby being comparatively cheap), and potentially empowers developers to work on architectures that they might not otherwise be able to \cite{eiras2024near}. This might also lead to the recognition of bugs, improving safety. A correlative argument is that open-weighting “democratizes AI”, giving “more people influence over how AI is developed and used, and promoting the representation of more diverse interests and needs in the direction of the field” \cite[p. 10]{seger2023open}.

At the same time, open-weight models introduce and exacerbate a number of risks \cite[Section 3]{seger2023open}. First, while closed source models themselves are not currently as well secured as would be desirable, open-weight models have been shown to be extremely vulnerable to adversarial detuning \cite{qi2023finetuning}, allowing malicious actors to both disable safeguards and elicit dangerous capabilities \cite{davidevan}. Even if models were closed thereafter, making understanding of model internals more broadly available might give malicious actors a greater chance of finding exploits that work around safeguards. Second, while it some bugs may be relatively easy to fix, others might be neither easy to spot nor fix, leading to vulnerabilities proliferation. Even if patches are built, it is difficult to ensure that these will actually be applied. It can also be difficult to assign liability to such harms, making it harder to incentivise actors against committing them \cite{davidevan}. 

Such risks operate in tandem with open-source infrastructure. A malicious actor could exploit the resources on HuggingFace, for instance, by downloading an open-weight model and fine-tuning it on an offensive dataset by following the instructions online (see section 3.3). This did in fact happen in 2023 with GPT4chan, a model trained on racist hate speech which was downloaded nearly 1,500 times before it was taken down \cite{harris2023howto}. Such platforms can make it harder to defend powerful models against jail-breaking or misuse. Although platforms like HuggingFace have already made a number of statements on ethics and policy \cite{huggingface2023content}, it is relatively easy to find posts on getting around safeguards on datasets and models (see Section 4.3). Similar sites might remain zones for building dangerous models, and sharing means to elicit dangerous capabilities. Securing against this possibility is difficult: were HuggingFace to tighten their reporting requirements, users might always move elsewhere. 

\subsection{Governing the Proliferation Paradigm: Principles}

The preceding subsections have sought to set out the contours of the Proliferation paradigm. Although the word `proliferation' aims to emphasise the dissemination of capabilities across actors and models that characterises most of the increase in risk, other words might also describe it. `Sub-frontier', `underbelly', `distributed' or simply `emerging' all capture important features. A set of overlapping hypotheses about the future, the paradigm is necessarily uncertain, new technological pathways may emerge, and lines dividing the aforementioned technologies may alter over time. `SHADOW' is not the final word on the AI proliferation, but a tool for beginning a conversation.

All SHADOW technologies weaken the assumptions of the 'Big Compute' paradigm, including governance strategies like compute governance. This creates three main disadvantages for governance mechanisms aiming to mitigate risks. 

First, it substantially reduces capability \textit{visibility}. Though it might seem like open-weighting models would increase visibility, doing so publicly means that it is very hard to keep track of which other actors might have access to those capabilities. Similarly, the lower the requirements for accessing sufficient compute, or capability keys necessary to elicit dangerous capabilities, the easier it will be to perpetrate harms. Paradoxically, the openness of benevolent or neutral actors enables secrecy to thrive. 

Second, the massively increased target site of the Proliferation paradigm would make policy far more difficult to \textit{enforce}. One axis here is model relevance. In an augmented-model paradigm, old models can be augmented to create powerful capabilities, just as compute sold to hidden actors years ago or latent in smart phones can be used to train dangerous systems. For models, the analogy of a `frontier' serves poorly; better to think of a `high water mark', with models below the surface remaining potentially (and dangerously) in play. The second axis is jurisdictional. Hidden models may operate across jurisdictions; information about capabilities augmentation or model weights is inherently international. Whilst \textit{compute} might have a centralised supply chain \cite{sastry2024computing}, \textit{capabilities} might not. Recognising and coordinating these is necessarily a global project, that will need to take into account global perspectives and various jurisdictions. 

Third, in a Proliferation paradigm, actions may be \textit{non-reversible}. Whereas a company can release access to a closed-source model and then restrict it, in open-weighting, as Seger et al. (2023) point out, “there are no take-backs" \cite{seger2023open}. The same goes for publishing capability keys online, or sharing highly efficient architectures. While threats of `regime lock-in' around AI governance should be taken seriously \cite{clark2023375}, the term might apply equally to the technological paradigm these new developments create, for better or for worse. 

Such technological affordances pose daunting challenges for risk-mitigation. The next section proposes three sets of strategies for meeting them effectively, foregrounding an ethical perspective, and making clear where more research will be required. 

\section{The Proliferation Paradigm: Towards Responsible Governance}

The reader should now have an understanding of the Proliferation paradigm and the challenges it presents to the assumptions of `Big Compute'. This raises the question: if not compute, what aspects of the AI ecosystem should be governed to mitigate risks from AI risks effectively? How should these governance mechanisms be calibrated to ensure that this respects the values and principles of a broadly liberal, democratic society?

To answer this question, this paper returns to the concept of the ‘AI triad’ mentioned in Section 2.2 \cite{buchanan2020triad}. If not ‘Big Compute’, it argues, there are three main components of AI that governance can usefully target: \textit{algorithms} (the weights and architectures required to build dangerously capable systems), \textit{decentralized compute} (the facilities for actors to access compute anonymously or bypass regulation) and \textit{dangerous inputs} (the information required to elicit dangerous capabilities without additional compute, or bypass safeguards in powerful models). In each domain, a subsection sets out the key ethical trade-offs risk-mitigation strategies have to face, sets out promising strategies for mitigating them, and clearly stipulating the directions required for further research. \newline

\subsection{Governing Algorithms: Towards Responsible Access Policies}

A prevalent concern about both AI technologies and AI governance is that it might limit the number of actors capable of building AI models, keeping power centralised in the hands of a few \cite{dafoe2018ai, seger2023democratising, eiras2024near}. This has attracted criticism from contemporary ethico-political philosophers. In `Toward a Theory of Justice for Artificial Intelligence’, for example, Iason Gabriel argues (following Rawls) that any just development of AI as a component of a broader sociotechnical social system should emphasise “fair equality of opportunity” \cite[p. 224]{gabriel2022toward}.

Under a Rawlsian framework of distributive justice, Gabriel might argue, the proliferation of algorithmic technologies that characterise AI proliferation are preferable to those of `Big Compute’. As Section 3 showed, small models may be easier for smaller companies and even individuals to train and run how they wish; decentralized compute may better support individual actors than traditional cloud provisions; and freely-published augmented and open-weight models provide capabilities to all, particularly those parties that would not be able to train models themselves \cite{eiras2024near}. As Seger et al. (2023) point out in their useful study of the topic, proponents of technologies like open-sourcing can, and frequently do, refer to such a broadening of the pool of people capable of building models is akin to `democratising' AI \cite{seger2023open}. 

A fundamental misunderstanding is worth addressing before the virtues of this argument is addressed. Although Seger et al.'s piece on the use of `democratisation' is a valuable review of the literature, it might permit others to be careless in the use of of the word `democracy'. `Democracy' derives laudatory weight from its association with a decision process undertaken by groups of people, not the distribution of a population's access to tools. One can talk meaningfully about `democratising' an AI company board, or `democratisation' of AI that is used to support democratic governance processes (in the same way that one might talk of `militarisation' of AI) but not the tools it provides. Tools can be made available to larger populations in lieu or in spite of democratic processes those populations have access to. Similarly populations abiding by democratic processes can, and may well choose to, restrict access to powerful tools, especially if they were worried about technologies undermining democratic processes (see \cite{kreps2023ai, manheim2019artificial, bontridder2021role}).

The extent to which the AI proliferation would increase the number of people in a population able to create capable models should not be pushed too far either. As of 2022, only 60\% of the world's population had access to internet \cite{owid-internet}; perhaps only as many as 28 million identify as software developers (about the population of Madagascar) \cite{statista2022}. Open-source communities are also not greatly diverse themselves: studies show that they are predominantly white, university educated, and male at present \cite{maslej2023report}. Third, the marginal benefit to disempowered communities may be slim: as Seger et al. point out, AI models used (for instance) in drug discovery may still benefit a vast majority, whether or not the users themselves are capable of operating the model \cite[p. 28]{seger2023democratising}. At the same time, as Sections 2.1 and 3 noted, these systems also bear serious risks which may well effect many. Some of these, like bioweapons, may distribute their threats more evenly; others, like cyberphising, scam and disinformation attacks, might actively target less empowered groups \cite{brundage2018malicious}. A Rawlsian like Gabriel may, in other words, choose to centralise control of AI in trusted parties, and focus on distributing the benefits through economic policies like taxation and universal basic income \cite{bruun2018artificial}. 

At the same time, however, it surely makes sense to broaden the pool of those capable of building models \textit{if this can be done safely}, and without empowering malicious actors. This seems to pose a tension one might think of as an `access-security trade-off’. How to govern the publication of algorithmic information so as to provide as many capabilities as possible to agents and services which will create benefits, whilst restricting as many capabilities as possible from actors and practises which will cause harms? A key tool here is `structured access', a feature of AI model interfaces which would allow “access to the tool, without giving them enough information to create a modified version” or learn how to bypass crucial safeguards, analogous “to how a keycard grants access to certain rooms of a building” \cite[p. 47]{shevlane2022structured}.

Two recent papers constitute state-of-the-art in structured access. One, by Irene Solaiman, envisages five `levels' of model release, from “fully open” to “fully closed”  \cite[p. 4]{solaiman2023gradient}. The other, by Bucknall and Trager, looks at the variables of AI access (metadata, inspection, modification, fine-tuning and sampling) that successive layers of an API (Application Programming Interface) might consist of. Bucknall and Trager's paper then goes a step further, arguing that “insufficient access to models frequently limits research, but that the access required varies greatly depending on the specific research area” \cite[p. 1]{bucknall2023structured}. Doing so offers a valuable contribution to work that aims to calibrate access to model internals responsibly. In the style of `Responsible Scaling Policies' \cite{rsp}, one could term these `Responsible Access Policies’. 

What questions would need to be answered in order to calibrate such access policies responsibly? First, policymakers need a more general view of the marginal uplift in benevolent actor capabilities \textit{per se} of varying degrees of model access. More research might conclude that having access to a model weights “significantly aids developers in creating models that are high-performing and specifically tailored to their use-case” \cite[p. 18]{eiras2024near} compared to no access, such that open-weighting has significant economic benefits to a number of actors compared to closed.  Though it seems unlikely, it might also be that open-weighting “particularly helps cater to less well-resourced languages, domains, and downstream tasks” (p. 18) compared to closed-models operated by highly-paid teams aiming to do just this \cite{geminiteam2024gemini}. Cynically, one might agree with Pilz, Heim and Brown's suspicion that the “leader advantage" of large models is likely to restrict beneficent outputs to relatively few big leaders, even as capabilities proliferate \cite{pilz2023increased}. Yet empirical studies as to the net marginal capabilities of actors using open-source as opposed to closed-source technologies, or economic projections of how an open-source might augment the long-term distribution of wealth in an economy, should still be a priority. Overlooking rigorous empirical research risks compromising the integrity of movements like open-sourcing, and invites over-regulation.

Second, policymakers need a more general view of the net marginal uplift in \textit{malicious} actor capabilities, under different technological assumptions. Some threats may be scaled out (be low \textit{net}): for instance, if a model could be shown to make cyberattacks easier to perpetrate, but also easier to defend, then it might represent less of a risk to open-weight \cite{garfinkel2021does}. Some threats might not be marginal: releasing access to model weights might make it substantially easier to perform valuable research and to clone the model without safeguards, whereas releasing access to fine-tuning might make it easier to perform valuable research, but without the same downsides. Finally, some threats might be different under different assumptions: if a powerful model is large, but malicious actors could easily gather the compute required to detune it anonymously, or it is relatively easy to augment, then one should be more cautious about open-weighting it. The feasibility of addressing risks should be kept in mind throughout: risks that are easier to defend against, like societal disruption, might be addressed anticipatorily through social policy \cite{bernardi2024societal}.

In order to responsibly calibrate structured access policies, policymakers need a view of the net marginal uplifts in both malicious and benevolent actor capabilities under different technological assumptions before they can make any meaningful decisions. The work required, even to inform rough estimates, is considerable: Bucknall and Trager's paper should set the standard for papers which perform new empirical research and summarise existing research and expert opinions for policymakers to map this rapidly evolving landscape. An ‘optimal’ calibration here should not just look for diminishing returns (that is, the point at which increasing the amount of access fails to increase capacity to do beneficial work but creates new risks) but a dynamic process which involves a wide variety of stakeholders, particularly those representing no-internet or no-code communities, to communicate about and converge on the value structures that they would wish to embody \cite{gordon2022mapping}. More guidance on how to elicit these values is suggested in the next section.

\subsection{Governing Decentralized Compute: Towards Privacy-Preserving Oversight}

A common refrain in AI ethics literature is that AI technologies and AI governance should respect individual privacy; specifically, individual's right to access and build AI systems without government oversight \cite{dafoe2018ai}. Pursuing Rawl’s “concern with the ability of citizens to pursue a conception of the good life that is free from unwarranted interference”, for instance, Iason Gabriel cites Andrei Marmor’s argument that this should include the right to privacy on the grounds that it is intimately connected to well-being \cite[p. 224]{gabriel2022toward}. Such a right, Marmor argues, is “violated when somebody manipulates, without adequate justification, the relevant environment in ways that significantly diminish your ability to control what aspects of yourself you reveal to others” (ibid).  

As Section 1.2 mentioned, the ‘Big Compute’ paradigm provided a number of advantages for governance mechanisms that would defend the privacy of actors aspiring to access compute. The set of relevant companies (or ‘hyperscalers’) capable of providing or affording sufficient compute to elicit dangerous capabilities was assumed to be relatively small and high profile. A Know-Your-Customer scheme of the sort proposed by Heim and Egan (2024) in the style of financial markets was consequently unlikely to impose a regulatory burden on smaller actors \cite[p. 7]{egan2023oversight}. As Sastry et al. (2024) note, compute has further potential to be a uniquely impersonal site for governance \cite{sastry2024computing}. Using privacy-preserving practises like workload monitoring, regulation need not require any information as to the function the compute would be used for, thereby protecting confidential company strategies or IP \cite[p. 86]{sastry2024computing}. 

A proliferation paradigm would upend many of these assumptions. The set of relevant actors is vast, including not only many companies and jurisdictions but potentially individuals; these individuals might be unwilling, or unable, to comply with the transparency burdens required of larger companies; and privacy-preserving oversight measures like work-load monitoring might not translate well to a paradigm of smaller models. At the same time, entering a Proliferation paradigm does not diminish the importance of regulating compute: if anything, by lending greater importance to small models, it increases it. Small models can be trained on widely available quantities of compute, and open-weighted models can be downloaded and adjustment on decentralized platforms. At the same time, as Sastry et al. suggest, “increasing visibility into AI-relevant computation could carry significant risks to privacy and civil liberties” \cite[p. 52]{sastry2024computing}. The question then is what level of privacy should be protected, for which actors, under what technological and societal circumstances?

One way to do this, following the previous section, would be to perform empirical research to estimate the net marginal uplift in malicious and benevolent actor capabilities from various levels of \textit{compute} access, and use it to inform regulatory policy. Empirically founded views here will be crucial to ensure that this does not become standard practise for malicious actors, a `black market for compute’. Keeping track of these datapoints also seems valuable in general: there may be models in the future that would be reasonably safe assuming very widely available levels of compute (e.g. desk top computers) but quickly become unsafe when that level is raised by several orders of magnitude (e.g. by using a compute cluster accessed off a compute market to adversarially detune them). On the other hand, if empirical research showed that there was no real uplift in risks from the level of compute generally provided on decentralized platforms (as may be the case at present) then these platforms should be protected from overregulation.

It seems likely, however, that the Proliferation paradigm would involve decentralized compute platforms which \textit{are} capable of creating some net uplift in risks. One way to regulate here would be to follow Heim and Egan's model of `Big Compute' in setting thresholds for the ability of unregistered anonymous users to interact with the system \cite{egan2023oversight}. For instance, actors can read posts on HuggingFace, but they cannot post anything themselves or download model weights without creating an account on the platform. Gensyn might, under certain circumstances, implement a similar policy of responsible access: actors could use the account to obtain a certain amount of compute, but beyond a certain threshold, would have to give up personal information or indicate the purpose of their usage. Those resisting these policies or failing to update their thresholds in line with changing technological environments might be shut down, following Sastry et al. (2024), in a manner akin to how “digital services are shut down for legal violations, such as hosting illegal online drug markets” \cite[p. 57]{sastry2024computing}.

This threshold would and should be highly contested. Five research questions can help policymakers estimate the ethical trade-offs. The first would be to ask who uses decentralized compute. To what extent are decentralized marketplaces used, or vulnerable to being used, by international actors with VPNs seeking to evade either national AI compute restrictions, or restrictions on the export of chips to that area? Second would be to ask how users of decentralized market places do in fact use them. Are they often used to train AI systems, or for other purposes (building video game simulation graphics, training climate simulation models)? Third, how much do such actors value the anonymity decentralized marketplaces provide: are they, for instance, attracted by cheaper price, or short-term contracts instead? Fourth, what sort of privacy do these actors require: are clients willing to share their name and details, but not the purpose of their compute usage, or vice versa? 

Fifth, assuming that these parties do value privacy, how persuasive are their reasons for doing so? As Helen Nissenbaum has argued, privacy norms are greatly informed by social practises: norms in one field are often translated into other, particularly into contemporary digital spheres, without a rigorous exposition of the purpose of these practises \cite{nissenbaum2004privacy}. To be clear, these reasons for privacy might be hugely important. If so, this research should usefully bolster the tendency to impose less rigorous anti-privacy regulation. On the other hand, other reasons to seek privacy might be less convincing (`the NSA is spying on me’) or tractable through other means (`Cloud compute know-your-customer requirements are far too tiresome for me to fill out’). Graphing and interpolating these unknowns could not only help third party marketplaces understand the needs of their consumers, but help to estimate the actual harm of enforcing anti-privacy measures on these groups. 

Of course, if the compute requirements to elicit dangerous capabilities were more minimal, but still within the domain of third-party compute providers, then the gradated access policy might not be secure enough, nor basic Know-Your-Customer policies sufficiently robust against malicious actors. Research into privacy preserving monitoring strategies, such as those proposed hypothetically by Sastry et al. \cite{sastry2024computing}, might then be a key priority. If these were difficult to implement, mitigating the number of decentralized markets \cite{mercille2019market}, the number of AI chips \cite{donnelly2023semiconductor}, or implementing on-chip workload tracking \cite{aarne2024secure} might all be tractable suggestions for preserving safety, though the significance level for empirical evidence for the existence of such threats (as opposed to e.g. regulatory capture by large AI firms) might have to be considerable. 

There remains, however, the `Vulnerable World’ \cite{bostrom2019vulnerable} scenario in which AI capabilities are so easy to elicit using augmented models, or small models, or even large models operated across distributed architectures, that any sort of governance faces the colossal and impossible task of regulating almost all devices with the level of compute in a smart phone. Under such scenarios, among others, the role of scientific information might be critical. This is the topic of the next section. 

\subsection{Governing Ideas: Towards Responsible Information Security}

Some scholars have contended that, in an ideal world, many people would have access not only to algorithms and the compute necessary to operate them, but to the information to design and elicit capabilities from them. In ‘A Human Rights-Based Approach to Responsible AI’ (2022), for instance, Prabhakaranan et al. argue that everyone has a right to “share in scientific advancement", where "science" describes "(1) knowledge, (2) the application of that knowledge, and (3) the method of the knowledge production." \cite[p.9]{prabhakaran2022human}, citing citing the Universal Human Rights Declaration (UHRD, Article 27). The right to share scientific information under freedom of speech (as protected, for instance, by the First Amendment) is debated by a number of legal scholars in the US \cite{robertson1977scientist, goldberg1979constitutional} and European \cite{verschraegen2018regulating}. 

There are cases where increasing public access to information about AI systems should be a priority. Patches resolving model bugs should proliferate; information about how to make models cheaper or more energy efficient might help to lower the customer and environmental costs. But clearly not \textit{all} information shared about AI systems is beneficial. One wouldn't want nuclear capability keys shared widely; similarly, model weights might deliver powerful capabilities to malicious actors; capability keys circulated online technique shared publicly compromise the safety of a powerful model.\footnote{Some of these risks might be able to solved by in a centralized manner (e.g. patching a centralised small model against jailbreaks); others might not, and if models are open-weight and downloadable, patches might proliferate, creating irreversible risks.} Taken literally, the UHRD is absurd: companies routinely patent and protect scientific IP, require workers in critical labs to sign NDAs, and designate issues as confidential. Yet the underlying challenge remains: what steps should individuals, platforms and research communities take to create principles for sharing scientific information pertaining to potentially dangerous AI capabilities that maximise the benefits from this communication, whilst minimising the risks?

The first step towards responsible information security requires identifying information with potential hazards. This bears strong resemblances to "infohazards" described in biotechnological literature (e.g. \cite{lewis2019information}). For AI, these might fall into two categories of `inputs'. The first can be thought of as \textit{jail-breaks} which bypass safeguards. This might include prompt-based jail-breaking strategies or anti-safeguard synthetic data for adversarial fine tuning (see 3.3). The second could be thought of as \textit{capability keys} which elicit unexpected capabilities from models without involving substantially more compute. This might include the weights of dangerously capable models (see 3.5), powerful pruning techniques (3.1), powerful fine-tuning datasets or prompt-scaffolding arrangements for agentic models (3.3), or efficient distributed architectures (3.4).   

While such evaluation might usefully borrow methodology from responsible access policies around algorithms, it is exceptionally difficult to ensure that there will be no unlocks or capabilty keys for LLMs. LLM outputs can be highly unpredictable \cite{glitch}, and the science for evaluating them is still nascent \cite{anwar2024foundational}, making it possible that vulnerabilities will be missed. Moreover, information security faces the substantially greater challenge of having to encapsulate many axes of communication, rather than just one (model release). These challenges can be broken down into two sorts: \textit{closed-group} policy related to groups like labs and companies developing new technologies, and \textit{open-group} policy related to platforms like HuggingFace and LAION sharing this information.  

Closed-group information communication poses one set of challenges. Workforces are constantly evolving; the information flow needing to be evaluated constantly is great; individual reputation might be involved with decisions about whether to share a new capability unlock. As a recent RAND paper on securing model weights observed of an AI company, key stakeholders may also disagree over key variables \cite[p. 32]{nevo2024securing}. It may be that substantial empirical research is required before stakeholders can reach agreements here. Given the fast nature of AI development, this may not be feasible. Yet it is precisely this that makes developing and articulating responsible information communication policies a priority. In a landscape of many unknowns, having robust norms and established practises for identifying which information is and isn't harmful, and how it should be communicated, is all the more crucial. 

Take, as a case study, the 2022 blog post setting out the ‘infohazard policy’ of AI company Conjecture. Despite setting out how various secrecy levels should relate to each other, the policy framework fails to  describe how they should relate to technologies, a conversation which might benefit other actors more significantly.
It offers no concrete analysis of why certain information falls into which bucket beyond the discretion of the ‘appointed infohazard coordinator’ \cite{conjecture}. This seems unlikely to scale effectively, and proposes no accountability on those infohazard coordinators in the event of an error. Yet given the lack of precedent or common practise here, the naivety is understandable. No other companies have released explicit infohazard policies, or commitments to mitigate infohazards. Though these are unlikely to be sufficiently rigorous, making commitments to set out explicit infohazard policies would be a start, and allow third parties and the public to compare the stringency of individual companies, how companies accountable for any leaks, and track how these policies evolved in line with responsible scaling, access, and security policies. Frameworks like the National Institute of Standards and Technology's information security framework \cite{NIST}, or the EU's General Data Protection Regulation required Data Protection Impact Assessments \cite{10.1145/3339252.3340516} might provide initial starting points. 

A prevailing concern here however is how to secure these models weights once these striations are decided. In a recent report setting out approximately 38 attack vectors from different capabilities of malicious actors, Nevo et al. (2024) note that securing model weights against the most capable actors (i.e. highly motivated states) is currently not possible, and there is substantial debate about how to achieve security against very capable actors (i.e. state-sponsored groups) \cite{nevo2024securing}. Securing weights may become harder as frontier models develop more powerful cybercapabilities \cite{garfinkel2021does}; information about capability keys might be far easier to elicit through Nevo et al.'s proposed attack vectors, especially if it isn't immediately identifiable as sensitive. 

\textit{Open-group} communication norms present a subtly different set of challenges. A key risk here is that a neutral or malicious actor shares weights or jailbreaks on a public site, which then proliferate. On the one hand, machine learning information platforms like EleutherAI Discord, HuggingFace, EA Forum, LessWrong, Alignment Forum and even more general platforms like Facebook and X might be usefully served by modelling infohazard policies on those of AI companies (Meta and X could use policies already developed in-house, for instance) to identify and take down content before it proliferates online. On the other hand, creating an information security regime that could realistically account for these practises is a serious challenge. Content moderation on large platforms has been viciously contested, reluctantly deployed, and ineffective \cite{goldstein2023understanding, nytimes}. Content policy already in action deserves further examination, but clearly displays flaws. For instance, HuggingFace suggests that it will downgrade, privatise or disable the visibility of ML artefacts on its platform, and asserts that it does not tolerate “code that is designed to disrupt, damage or gain unauthorized access to a computer system or device” or “proxies that are primarily designed to bypass restrictions imposed by the original service provider”, among other technical restrictions \cite{huggingface2023content}. In practise, however, it is trivially easy to find material that skirts the lines of these restrictions (figure 1). 

\begin{figure}
    \centering
    \includegraphics[width=1\linewidth]{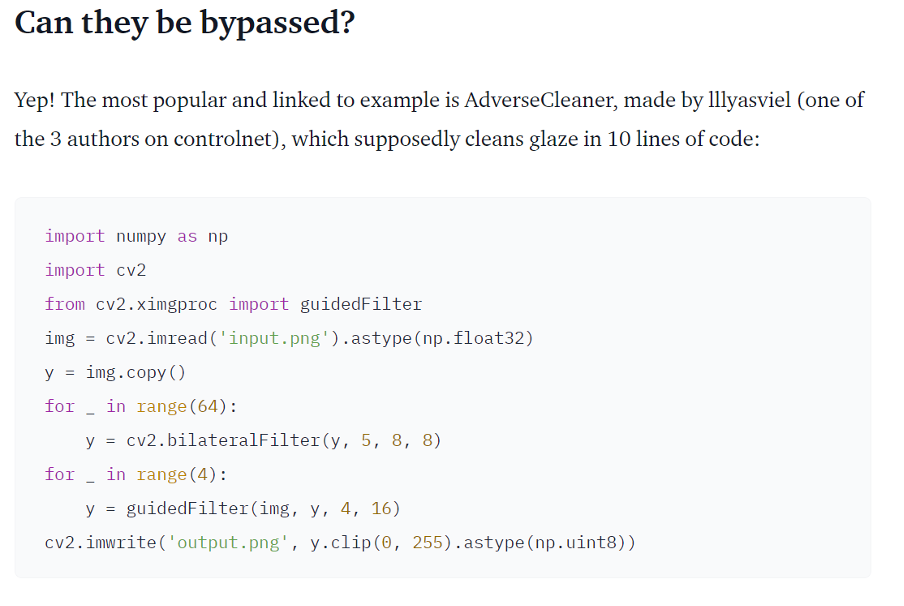}
    \caption{It took about 2 minutes to find a HuggingFace post describing how to bypass Glaze, a measure designed to protect art against AI mimicry. \cite{glaze2024}}
    \label{fig:enter-label}
\end{figure}

As with decentralized compute, governance actors may struggle to contend with established norms of free information sharing developed in one paradigm which translate poorly to proliferation. Even then, analogies from biotechnology suggest that content like papers may be published with dangerous dual-use capabilities \cite{imai2012experimental, noyce2018construction, lewis2019information}, even after the infohazard awareness in that field was relatively mature \cite{atlas2005}. Such policy might still be worthwhile: in information epidemiology, restricting the transmission of information from one site to another even slightly (say, each person telling 1 other rather than 5) can have significant compounding effects. Mitigating the spread of a leak – even just to give governance mitigations days or hours of head start – can mean the difference between deploying patches and serious harms. 

The bottom line here is that an AI paradigm in which powerful models are jailbroken or elicited by capability keys would be extremely difficult to govern responsibly, even assuming advanced information policy and security. Once unsecured, this information is likely to proliferate, potentially creating irreversible risks. The low feasibility of this governance approach to mitigating harms would seem a significant reason to be cautious about model releases, especially open-weight models, and to prioritise model robustness against jailbreaks, for instance via adversarial testing \cite{ganguli2022red}. At the extreme end, specific research objectives with clear dual use capabilities to avoid governance, like Deepmind’s DiPaCo research stream, should be carefully evaluated as to the extent they might enable malicious actors before research is made public, or should be proposed only alongside strategies for governing the risks. Sometimes the most secure way to deal with a possible information hazard is not to put funds towards developing it at all.

\subsection{Governing Proliferation: Key Principles}

Each governance strategy---responsible access policies, privacy-preserving oversight, and strong information security measures---fundamentally requires three things. First, an empirically rigorous basis for estimating the net marginal uplift in malicious and benevolent actor capabilities from different levels of access (to compute, algorithms or information) under different technological assumptions. Second, a well-substantiated critical literature reflecting stakeholder values that can be used to navigate the necessary trade-offs, even as the technology emerges. The third requirement is security. Unless sensitive information like model weights and capability keys can be effectively secured, governance promises (like responsible access policies) will be futile. Future research in all three domains is necessary and urgent if responsible governance strategies are to develop at the speed of developing technology.

Two further points might usefully orientate future research. First, just as policy designed for one paradigm might not be suitable for another, ethical norms that earned respect in one technological paradigm shouldn't be assumed to hold value in another. Free communication strategies designed around harmless scientific factoids cannot apply to capability keys or jail-breaks, nor privacy ethics for agents training AI models which could hack private infrastructure. Two implications stand out. If decision makers wish to champion values that acquired evaluative weight before AI proliferation, they should do so conditional on recognition of their cost. Alternatively, if some values were assumed to be absolute, then decision makers should be careful about funding particular technology that enables behaviours antithetical to those values. Further research into how the risk landscape affected by technological developments informs, and alters, ethical norms will be necessary to avoid undesirable ethical drift.

Second, even given empirically-driven estimates for relevant uplifts, and a clear view of societal values, policymakers still face the problem of inductive risk \cite{douglas2000inductive}. Should models be assumed safe until proven unsafe (loosely speaking, the `accelerationist' position), or vice versa? A useful heuristic here is to consider the lens of \textit{regime lock-in}, and operate under the principle of `accelerate when actions are reversible, and deccelerate or even pause where irreversible harms may arise'. Such a framework might obtain the best of both worlds, might be bullish on open, free API access for powerful models; it might be bearish on publication, capability keys, and prioritise cyber- and information-security. Fine-graining what such a policy strategy might look like in practise would be a valuable line for future research. 

\section{Conclusion}

\begin{quote}
    \textit{You cannot [do anything to] control a technology which gets more than a hundred times cheaper to do in half a decade. Not a thing!} 
    \\ \tabto{5mm} - Jack Clark, GPT2, Five Years On \cite{clark2023375}
\end{quote}

As well as benefits, dangerously capable AI models already present significant risks and challenges for governance. The technological pathways that constitute AI proliferation are already coming into existence, will continue to develop, pose substantial risks, and are both less visible to regulators and harder to enforce against. Several strategies for mitigating these risks look promising, but it will require considerable further research to ensure they are calibrated, and defended, both effectively and ethically. 

How should policymakers conceptualise the scope of this new challenge, and their individual agency within it? This paper has taken, for an organising metaphor for AI, the nuclear bomb. Yet a tightly centralised, expert-driven, vastly expensive national project with a binary output (a society either `has' atomic weapons, or it doesn't) is far more reminiscent of `Big Compute' than AI proliferation. 
A Proliferation paradigm would involve vastly greater numbers of actors across states, industries and publics; complex interactions between multiple technological pathways; and a more gradual diffusion of risky capabilities from major labs into the broader population than any `singularity'. If the new paradigm suits a metaphor, it is closer to climate change: self-interested actors acting as individuals, creating compound effects that gradually introduce greater risks into the ecosystem, affecting sudden shifts that require global action to mitigate.

This view sounds pessimistic. Yet an AI proliferation paradigm is unlikely to stumble on a “vulnerable world" of the sort Segrè's bomb suggests \cite{bostrom2019vulnerable}. Jack Clark's thesis, above, is overly deterministic: if high-risk capabilities arrive without security and safeguards, it will be due to a panoply of individual decisions. Developers, and the public, have real agency here. The benefits of AI proliferation are immense, and if society is capable of achieving them without irresponsible risks, they should try to do so. At the same time, if risks cannot be mitigated effectively, or mitigated ethically, then pausing, rerouting funding away from high-risk strategies, and steering towards reversible experimentation, should all be real considerations. Without sufficient safeguards or societal resilience, societies that hurtle towards AI proliferation might do so at their peril.

\newpage
\printbibliography

@book{rhodes2012making,
  title={The Making of the Atomic Bomb},
  author={Rhodes, Richard},
  year={2012},
  publisher={Simon and Schuster}
}

@article{bostrom2019vulnerable,
  title={The vulnerable world hypothesis},
  author={Bostrom, Nick},
  journal={Global Policy},
  volume={10},
  number={4},
  pages={455--476},
  year={2019},
  publisher={Wiley Online Library}
}

@article{ho2015typology,
  title={A typology of technological change: Technological paradigm theory with validation and generalization from case studies},
  author={Ho, Jonathan C and Lee, Chung-Shing},
  journal={Technological Forecasting and Social Change},
  volume={97},
  pages={128--139},
  year={2015},
  publisher={Elsevier}
}

@article{dosi1982technological,
  title={Technological paradigms and technological trajectories: a suggested interpretation of the determinants and directions of technical change},
  author={Dosi, Giovanni},
  journal={Research policy},
  volume={11},
  number={3},
  pages={147--162},
  year={1982},
  publisher={Elsevier}
}

@book{kuhn1997structure,
  title={The structure of scientific revolutions},
  author={Kuhn, Thomas S},
  volume={962},
  year={1997},
  publisher={University of Chicago press Chicago}
}

@article{seger2023open,
  title={Open-sourcing highly capable foundation models: An evaluation of risks, benefits, and alternative methods for pursuing open-source objectives},
  author={Seger, Elizabeth and Dreksler, Noemi and Moulange, Richard and Dardaman, Emily and Schuett, Jonas and Wei, K and Winter, Christoph and Arnold, Mackenzie and h{\'E}igeartaigh, Se{\'a}n {\'O} and Korinek, Anton and others},
  journal={arXiv preprint arXiv:2311.09227},
  year={2023}
}

@article{heim2024training,
  title={Training Compute Thresholds: Features and Functions in AI Governance},
  author={Heim, Lennart},
  journal={arXiv preprint arXiv:2405.10799},
  year={2024}
}

@article{sastry2024computing,
  title={Computing Power and the Governance of Artificial Intelligence},
  author={Sastry, Girish and Heim, Lennart and Belfield, Haydn and Anderljung, Markus and Brundage, Miles and Hazell, Julian and O'Keefe, Cullen and Hadfield, Gillian K and Ngo, Richard and Pilz, Konstantin and others},
  journal={arXiv preprint arXiv:2402.08797},
  year={2024}
}

@article{anderljung2023frontier,
  title={Frontier AI regulation: Managing emerging risks to public safety},
  author={Anderljung, Markus and Barnhart, Joslyn and Leung, Jade and Korinek, Anton and O'Keefe, Cullen and Whittlestone, Jess and Avin, Shahar and Brundage, Miles and Bullock, Justin and Cass-Beggs, Duncan and others},
  journal={arXiv preprint arXiv:2307.03718},
  year={2023}
}

@techreport{anthropic2023scaling,
    author = {Anthropic},
    title = {Anthropic's Responsible Scaling Policy},
    institution = {Anthropic},
    year = {2023}
}

@proceedings{EUAIAct,
    key = {XXX},
    title = {EU AI Act},
    year = {2021}
}

@article{hoffmann2022training,
  title={Training compute-optimal large language models},
  author={Hoffmann, Jordan and Borgeaud, Sebastian and Mensch, Arthur and Buchatskaya, Elena and Cai, Trevor and Rutherford, Eliza and Casas, Diego de Las and Hendricks, Lisa Anne and Welbl, Johannes and Clark, Aidan and others},
  journal={arXiv preprint arXiv:2203.15556},
  year={2022}
}

@inproceedings{sevilla2022compute,
  title={Compute trends across three eras of machine learning},
  author={Sevilla, Jaime and Heim, Lennart and Ho, Anson and Besiroglu, Tamay and Hobbhahn, Marius and Villalobos, Pablo},
  booktitle={2022 International Joint Conference on Neural Networks (IJCNN)},
  pages={1--8},
  year={2022},
  organization={IEEE}
}

@article{abdin2024phi,
  title={Phi-3 technical report: A highly capable language model locally on your phone},
  author={Abdin, Marah and Jacobs, Sam Ade and Awan, Ammar Ahmad and Aneja, Jyoti and Awadallah, Ahmed and Awadalla, Hany and Bach, Nguyen and Bahree, Amit and Bakhtiari, Arash and Behl, Harkirat and others},
  journal={arXiv preprint arXiv:2404.14219},
  year={2024}
}

@article{douillard2024dipaco,
  title={DiPaCo: Distributed Path Composition},
  author={Douillard, Arthur and Feng, Qixuan and Rusu, Andrei A and Kuncoro, Adhiguna and Donchev, Yani and Chhaparia, Rachita and Gog, Ionel and Ranzato, Marc'Aurelio and Shen, Jiajun and Szlam, Arthur},
  journal={arXiv preprint arXiv:2403.10616},
  year={2024}
}

@article{bai2023qwen,
  title={Qwen technical report},
  author={Bai, Jinze and Bai, Shuai and Chu, Yunfei and Cui, Zeyu and Dang, Kai and Deng, Xiaodong and Fan, Yang and Ge, Wenbin and Han, Yu and Huang, Fei and others},
  journal={arXiv preprint arXiv:2309.16609},
  year={2023}
}

@misc{davidson2023ai,
      title={AI capabilities can be significantly improved without expensive retraining}, 
      author={Tom Davidson and Jean-Stanislas Denain and Pablo Villalobos and Guillem Bas},
      year={2023},
      eprint={2312.07413},
      archivePrefix={arXiv},
      primaryClass={cs.AI}
}

@article{pilz2023increased,
  title={Increased Compute Efficiency and the Diffusion of AI Capabilities},
  author={Pilz, Konstantin and Heim, Lennart and Brown, Nicholas},
  journal={arXiv preprint arXiv:2311.15377},
  year={2023}
}

@techreport{scharre2024future,
    author = {Scharre, Paul},
    title = {Future-Proofing Frontier AI Regulation},
    institution = {Center For National American Security},
    year = {2024}
}

@article{eiras2024near,
  title={Near to Mid-term Risks and Opportunities of Open Source Generative AI},
  author={Eiras, Francisco and Petrov, Aleksandar and Vidgen, Bertie and de Witt, Christian Schroeder and Pizzati, Fabio and Elkins, Katherine and Mukhopadhyay, Supratik and Bibi, Adel and Csaba, Botos and Steibel, Fabro and others},
  journal={arXiv preprint arXiv:2404.17047},
  year={2024}
}

@article{dafoe2024ai,
  title={AI Governance},
  author={Dafoe, Allan},
  journal={The Oxford Handbook of AI Governance},
  pages={21},
  year={2024},
  publisher={Oxford University Press}
}

@article{schneider2020ai,
  title={AI governance for businesses},
  author={Schneider, Johannes and Abraham, Rene and Meske, Christian and Brocke, Jan vom},
  journal={arXiv preprint arXiv:2011.10672},
  year={2020}
}

@article{taeihagh2021governance,
  title={Governance of artificial intelligence},
  author={Taeihagh, Araz},
  journal={Policy and society},
  volume={40},
  number={2},
  pages={137--157},
  year={2021},
  publisher={Oxford University Press}
}

@article{ho2023international,
  title={International institutions for advanced AI},
  author={Ho, Lewis and Barnhart, Joslyn and Trager, Robert and Bengio, Yoshua and Brundage, Miles and Carnegie, Allison and Chowdhury, Rumman and Dafoe, Allan and Hadfield, Gillian and Levi, Margaret and others},
  journal={arXiv preprint arXiv:2307.04699},
  year={2023}
}

@article{sandbrink2023artificial,
  title={Artificial intelligence and biological misuse: Differentiating risks of language models and biological design tools},
  author={Sandbrink, Jonas B},
  journal={arXiv preprint arXiv:2306.13952},
  year={2023}
}

@techreport{openai2024bio,
    author = {OpenAI},
    title = {Building an early warning system for LLM-aided biological threat creation},
    institution = {OpenAI},
    year = {2024)}
}

@article{mirsky2023threat,
  title={The threat of offensive ai to organizations},
  author={Mirsky, Yisroel and Demontis, Ambra and Kotak, Jaidip and Shankar, Ram and Gelei, Deng and Yang, Liu and Zhang, Xiangyu and Pintor, Maura and Lee, Wenke and Elovici, Yuval and others},
  journal={Computers and Security},
  volume={124},
  pages={103006},
  year={2023},
  publisher={Elsevier}
}

@article{malatji2024artificial,
  title={Artificial intelligence (AI) cybersecurity dimensions: a comprehensive framework for understanding adversarial and offensive AI},
  author={Malatji, Masike and Tolah, Alaa},
  journal={AI and Ethics},
  pages={1--28},
  year={2024},
  publisher={Springer}
}

@article{shevlane2023model,
  title={Model evaluation for extreme risks},
  author={Shevlane, Toby and Farquhar, Sebastian and Garfinkel, Ben and Phuong, Mary and Whittlestone, Jess and Leung, Jade and Kokotajlo, Daniel and Marchal, Nahema and Anderljung, Markus and Kolt, Noam and others},
  journal={arXiv preprint arXiv:2305.15324},
  year={2023}
}

@article{anwar2024foundational,
  title={Foundational challenges in assuring alignment and safety of large language models},
  author={Anwar, Usman and Saparov, Abulhair and Rando, Javier and Paleka, Daniel and Turpin, Miles and Hase, Peter and Lubana, Ekdeep Singh and Jenner, Erik and Casper, Stephen and Sourbut, Oliver and others},
  journal={arXiv preprint arXiv:2404.09932},
  year={2024}
}

@article{ji2023ai,
  title={Ai alignment: A comprehensive survey},
  author={Ji, Jiaming and Qiu, Tianyi and Chen, Boyuan and Zhang, Borong and Lou, Hantao and Wang, Kaile and Duan, Yawen and He, Zhonghao and Zhou, Jiayi and Zhang, Zhaowei and others},
  journal={arXiv preprint arXiv:2310.19852},
  year={2023}
}

@techreport{acemoglu2021harms,
  title={Harms of AI},
  author={Acemoglu, Daron},
  year={2021},
  institution={National Bureau of Economic Research}
}

@article{slee2020incompatible,
  title={The incompatible incentives of private-sector AI},
  author={Slee, Tom},
  journal={The Oxford Handbook of Ethics of AI},
  pages={106--123},
  year={2020},
  publisher={Oxford University Press Oxford}
}

@misc{moulange2023responsible,
      title={Towards Responsible Governance of Biological Design Tools}, 
      author={Richard Moulange and Max Langenkamp and Tessa Alexanian and Samuel Curtis and Morgan Livingston},
      year={2023},
      eprint={2311.15936},
      archivePrefix={arXiv},
      primaryClass={cs.CY}
}

@book{Veliz2020-VLIPIP,
	address = {London, UK},
	author = {Carissa V\'{e}liz},
	editor = {},
	publisher = {Penguin (Bantam Press)},
	title = {Privacy is Power},
	year = {2020}
}

@article{veliz2020data,
  title={Data, Privacy, and the Individual: A Report for the Center for the Governance of Change (https://philarchive.org/rec/VLIPM)},
  author={V{\'e}liz, Carissa},
  year={2020}
}

@article{richardson2021framework,
  title={A framework for fairness: A systematic review of existing fair ai solutions},
  author={Richardson, Brianna and Gilbert, Juan E},
  journal={arXiv preprint arXiv:2112.05700},
  year={2021}
}

@article{novelli2023accountability,
  title={Accountability in artificial intelligence: what it is and how it works},
  author={Novelli, Claudio and Taddeo, Mariarosaria and Floridi, Luciano},
  journal={AI and SOCIETY},
  pages={1--12},
  year={2023},
  publisher={Springer}
}

@inproceedings{seger2023democratising,
  title={Democratising AI: Multiple meanings, goals, and methods},
  author={Seger, Elizabeth and Ovadya, Aviv and Siddarth, Divya and Garfinkel, Ben and Dafoe, Allan},
  booktitle={Proceedings of the 2023 AAAI/ACM Conference on AI, Ethics, and Society},
  pages={715--722},
  year={2023}
}

@article{zeng2020artificial,
  title={Artificial intelligence and China's authoritarian governance},
  author={Zeng, Jinghan},
  journal={International Affairs},
  volume={96},
  number={6},
  pages={1441--1459},
  year={2020},
  publisher={Oxford University Press}
}

@article{dafoe2018ai,
  title={AI governance: a research agenda},
  author={Dafoe, Allan},
  journal={Governance of AI Program, Future of Humanity Institute, University of Oxford: Oxford, UK},
  volume={1442},
  pages={1443},
  year={2018}
}

@article{douglas2000inductive,
 ISSN = {00318248, 1539767X},
 URL = {http://www.jstor.org/stable/188707},
 author = {Heather Douglas},
 journal = {Philosophy of Science},
 number = {4},
 pages = {559--579},
 publisher = {[The University of Chicago Press, Philosophy of Science Association]},
 title = {Inductive Risk and Values in Science},
 urldate = {2024-06-04},
 volume = {67},
 year = {2000}
}

@misc{owen2024predictable,
      title={How predictable is language model benchmark performance?}, 
      author={David Owen},
      year={2024},
      eprint={2401.04757},
      archivePrefix={arXiv},
      primaryClass={cs.LG}
}

@article{panjwani2020study, 
	title={Study of Cloud Security in Hyper-scalers}, 
	DOI={https://doi.org/10.23919/indiacom49435.2020.9083727}, 
	author={Panjwani, Megha and De, Suman}, 
	year={2020},
 }

@article{wirtz2024ecosystem,
    author = {Wirtz, Bernd W. and Langer, Paul F. and Weyerer, Jan C.},
    title = {'An Ecosystem Framework of AI Governance'},
    journal = {The Oxford Handbook of AI Governance, Justin B. Bullock, and others (eds)},
    year = {2024}
}

@article{biden2023executive,
  title={Executive order on the safe, secure, and trustworthy development and use of artificial intelligence},
  author={Biden, Joseph R},
  year={2023}
}

@techreport{buchanan2020triad,
    author = {Buchanan, Ben},
    title = {"The AI Triad and What It Means for National Security Strategy"},
    institution = {Center for Security and Emerging Technology},
    year = {August 2020}
}

@misc{pilz2024increased,
      title={Increased Compute Efficiency and the Diffusion of AI Capabilities}, 
      author={Konstantin Pilz and Lennart Heim and Nicholas Brown},
      year={2024},
      eprint={2311.15377},
      archivePrefix={arXiv},
      primaryClass={cs.CY}
}

@misc{hernandez2020measuring,
      title={Measuring the Algorithmic Efficiency of Neural Networks}, 
      author={Danny Hernandez and Tom B. Brown},
      year={2020},
      eprint={2005.04305},
      archivePrefix={arXiv},
      primaryClass={cs.LG}
}

@misc{ho2024algorithmic,
      title={Algorithmic progress in language models}, 
      author={Anson Ho and Tamay Besiroglu and Ege Erdil and David Owen and Robi Rahman and Zifan Carl Guo and David Atkinson and Neil Thompson and Jaime Sevilla},
      year={2024},
      eprint={2403.05812},
      archivePrefix={arXiv},
      primaryClass={cs.CL}
}

@misc{gholami2021survey,
      title={A Survey of Quantization Methods for Efficient Neural Network Inference}, 
      author={Amir Gholami and Sehoon Kim and Zhen Dong and Zhewei Yao and Michael W. Mahoney and Kurt Keutzer},
      year={2021},
      eprint={2103.13630},
      archivePrefix={arXiv},
      primaryClass={cs.CV}
}

@misc{cheng2023survey,
      title={A Survey on Deep Neural Network Pruning-Taxonomy, Comparison, Analysis, and Recommendations}, 
      author={Hongrong Cheng and Miao Zhang and Javen Qinfeng Shi},
      year={2023},
      eprint={2308.06767},
      archivePrefix={arXiv},
      primaryClass={cs.LG}
}

@misc{kaleem2024comprehensive,
      title={A Comprehensive Review of Knowledge Distillation in Computer Vision}, 
      author={Sheikh Musa Kaleem and Tufail Rouf and Gousia Habib and Tausifa jan Saleem and Brejesh Lall},
      year={2024},
      eprint={2404.00936},
      archivePrefix={arXiv},
      primaryClass={cs.CV}
}

@misc{xu2023parameterefficient,
      title={Parameter-Efficient Fine-Tuning Methods for Pretrained Language Models: A Critical Review and Assessment}, 
      author={Lingling Xu and Haoran Xie and Si-Zhao Joe Qin and Xiaohui Tao and Fu Lee Wang},
      year={2023},
      eprint={2312.12148},
      archivePrefix={arXiv},
      primaryClass={cs.CL}
}

@misc{brown2020language,
      title={Language Models are Few-Shot Learners}, 
      author={Tom B. Brown and Benjamin Mann and Nick Ryder and Melanie Subbiah and Jared Kaplan and Prafulla Dhariwal and Arvind Neelakantan and Pranav Shyam and Girish Sastry and Amanda Askell and Sandhini Agarwal and Ariel Herbert-Voss and Gretchen Krueger and Tom Henighan and Rewon Child and Aditya Ramesh and Daniel M. Ziegler and Jeffrey Wu and Clemens Winter and Christopher Hesse and Mark Chen and Eric Sigler and Mateusz Litwin and Scott Gray and Benjamin Chess and Jack Clark and Christopher Berner and Sam McCandlish and Alec Radford and Ilya Sutskever and Dario Amodei},
      year={2020},
      eprint={2005.14165},
      archivePrefix={arXiv},
      primaryClass={cs.CL}
}

@misc{mehta2024openelm,
      title={OpenELM: An Efficient Language Model Family with Open Training and Inference Framework}, 
      author={Sachin Mehta and Mohammad Hossein Sekhavat and Qingqing Cao and Maxwell Horton and Yanzi Jin and Chenfan Sun and Iman Mirzadeh and Mahyar Najibi and Dmitry Belenko and Peter Zatloukal and Mohammad Rastegari},
      year={2024},
      eprint={2404.14619},
      archivePrefix={arXiv},
      primaryClass={cs.CL}
}

@misc{shazeer2017outrageously,
      title={Outrageously Large Neural Networks: The Sparsely-Gated Mixture-of-Experts Layer}, 
      author={Noam Shazeer and Azalia Mirhoseini and Krzysztof Maziarz and Andy Davis and Quoc Le and Geoffrey Hinton and Jeff Dean},
      year={2017},
      eprint={1701.06538},
      archivePrefix={arXiv},
      primaryClass={cs.LG}
}

@techreport{databricks2024dbrx,
    author = {Mosaic Research Team},
    title = {Introducing DBRX: A New State-of-the-Art Open LLM},
    institution = {Databricks},
    year = {March 27, 2024)}
}

@techreport{artic2024snowflake,
    author = {Snowflake AI Research},
    title = {Snowflake Arctic: The Best LLM for Enterprise AI — Efficiently Intelligent, Truly Open
},
    institution = {Snowflake},
    year = {April 24 2024}
}

@misc{liu2024kan,
      title={KAN: Kolmogorov-Arnold Networks}, 
      author={Ziming Liu and Yixuan Wang and Sachin Vaidya and Fabian Ruehle and James Halverson and Marin Soljačić and Thomas Y. Hou and Max Tegmark},
      year={2024},
      eprint={2404.19756},
      archivePrefix={arXiv},
      primaryClass={cs.LG}
}

@article{thawakar2024mobillama,
  title={MobiLlama: Towards Accurate and Lightweight Fully Transparent GPT},
  author={Thawakar, Omkar and Vayani, Ashmal and Khan, Salman and Cholakal, Hisham and Anwer, Rao M and Felsberg, Michael and Baldwin, Tim and Xing, Eric P and Khan, Fahad Shahbaz},
  journal={arXiv preprint arXiv:2402.16840},
  year={2024}
}

@techreport{lambert2024artic,
    author = {Lambert, Nathan},
    title = {Phi 3 and Arctic: Outlier LMs are hints},
    institution = {Interconnects.ai},
    year = {April 30 2024}
}

@techreport{IBM2024allam,
    author = {IBM},
    title = {Through partnership with IBM, Saudi Data and Artificial Intelligence Authority (SDAIA) launches a groundbreaking Arabic AI model to the Middle East},
    institution = {IBM},
    year = {May 21 2024}
}

@techreport{reuters2023falcon,
    author = {Barrington, Lisa},
    title = {Abu Dhabi makes its Falcon 40B AI model open sourc},
    institution = {Reuters},
    year = {May 25 2023}
}

@misc{brundage2018malicious,
      title={The Malicious Use of Artificial Intelligence: Forecasting, Prevention, and Mitigation}, 
      author={Miles Brundage and Shahar Avin and Jack Clark and Helen Toner and Peter Eckersley and Ben Garfinkel and Allan Dafoe and Paul Scharre and Thomas Zeitzoff and Bobby Filar and Hyrum Anderson and Heather Roff and Gregory C. Allen and Jacob Steinhardt and Carrick Flynn and Seán Ó hÉigeartaigh and Simon Beard and Haydn Belfield and Sebastian Farquhar and Clare Lyle and Rebecca Crootof and Owain Evans and Michael Page and Joanna Bryson and Roman Yampolskiy and Dario Amodei},
      year={2018},
      eprint={1802.07228},
      archivePrefix={arXiv},
      primaryClass={cs.AI}
}

@techreport{techopedia2024gpt2,
    author = {Keary, Tim},
    title = {GPT-2: The Strange Debacle Surrounding Mystery ‘OpenAI’ Chatbot},
    institution = {Techopedia},
    year = {11 May 2024}
}

@misc{kolt2024responsible,
      title={Responsible Reporting for Frontier AI Development}, 
      author={Noam Kolt and Markus Anderljung and Joslyn Barnhart and Asher Brass and Kevin Esvelt and Gillian K. Hadfield and Lennart Heim and Mikel Rodriguez and Jonas B. Sandbrink and Thomas Woodside},
      year={2024},
      eprint={2404.02675},
      archivePrefix={arXiv},
      primaryClass={cs.CY}
}

@misc{geminiteam2024gemini,
      title={Gemini 1.5: Unlocking multimodal understanding across millions of tokens of context}, 
      author={Gemini Team and Machel Reid and Nikolay Savinov and Denis Teplyashin and Dmitry and Lepikhin and Timothy Lillicrap and Jean-baptiste Alayrac and Radu Soricut and Angeliki Lazaridou and Orhan Firat and Julian Schrittwieser and Ioannis Antonoglou and Rohan Anil and Sebastian Borgeaud and Andrew Dai and Katie Millican and Ethan Dyer and Mia Glaese and Thibault Sottiaux and Benjamin Lee and Fabio Viola and Malcolm Reynolds and Yuanzhong Xu and James Molloy and Jilin Chen and Michael Isard and Paul Barham and Tom Hennigan and Ross McIlroy and Melvin Johnson and Johan Schalkwyk and Eli Collins and Eliza Rutherford and Erica Moreira and Kareem Ayoub and Megha Goel and Clemens Meyer and Gregory Thornton and Zhen Yang and Henryk Michalewski and Zaheer Abbas and Nathan Schucher and Ankesh Anand and Richard Ives and James Keeling and Karel Lenc and Salem Haykal and Siamak Shakeri and Pranav Shyam and Aakanksha Chowdhery and Roman Ring and Stephen Spencer and Eren Sezener and Luke Vilnis and Oscar Chang and Nobuyuki Morioka and George Tucker and Ce Zheng and Oliver Woodman and Nithya Attaluri and Tomas Kocisky and Evgenii Eltyshev and Xi Chen and Timothy Chung and Vittorio Selo and Siddhartha Brahma and Petko Georgiev and Ambrose Slone and Zhenkai Zhu and James Lottes and Siyuan Qiao and Ben Caine and Sebastian Riedel and Alex Tomala and Martin Chadwick and Juliette Love and Peter Choy and Sid Mittal and Neil Houlsby and Yunhao Tang and Matthew Lamm and Libin Bai and Qiao Zhang and Luheng He and Yong Cheng and Peter Humphreys and Yujia Li and Sergey Brin and Albin Cassirer and Yingjie Miao and Lukas Zilka and Taylor Tobin and Kelvin Xu and Lev Proleev and Daniel Sohn and Alberto Magni and Lisa Anne Hendricks and Isabel Gao and Santiago Ontanon and Oskar Bunyan and Nathan Byrd and Abhanshu Sharma and Biao Zhang and Mario Pinto and Rishika Sinha and Harsh Mehta and Dawei Jia and Sergi Caelles and Albert Webson and Alex Morris and Becca Roelofs and Yifan Ding and Robin Strudel and Xuehan Xiong and Marvin Ritter and Mostafa Dehghani and Rahma Chaabouni and Abhijit Karmarkar and Guangda Lai and Fabian Mentzer and Bibo Xu and YaGuang Li and Yujing Zhang and Tom Le Paine and Alex Goldin and Behnam Neyshabur and Kate Baumli and Anselm Levskaya and Michael Laskin and Wenhao Jia and Jack W. Rae and Kefan Xiao and Antoine He and Skye Giordano and Lakshman Yagati and Jean-Baptiste Lespiau and Paul Natsev and Sanjay Ganapathy and Fangyu Liu and Danilo Martins and Nanxin Chen and Yunhan Xu and Megan Barnes and Rhys May and Arpi Vezer and Junhyuk Oh and Ken Franko and Sophie Bridgers and Ruizhe Zhao and Boxi Wu and Basil Mustafa and Sean Sechrist and Emilio Parisotto and Thanumalayan Sankaranarayana Pillai and Chris Larkin and Chenjie Gu and Christina Sorokin and Maxim Krikun and Alexey Guseynov and Jessica Landon and Romina Datta and Alexander Pritzel and Phoebe Thacker and Fan Yang and Kevin Hui and Anja Hauth and Chih-Kuan Yeh and David Barker and Justin Mao-Jones and Sophia Austin and Hannah Sheahan and Parker Schuh and James Svensson and Rohan Jain and Vinay Ramasesh and Anton Briukhov and Da-Woon Chung and Tamara von Glehn and Christina Butterfield and Priya Jhakra and Matthew Wiethoff and Justin Frye and Jordan Grimstad and Beer Changpinyo and Charline Le Lan and Anna Bortsova and Yonghui Wu and Paul Voigtlaender and Tara Sainath and Shane Gu and Charlotte Smith and Will Hawkins and Kris Cao and James Besley and Srivatsan Srinivasan and Mark Omernick and Colin Gaffney and Gabriela Surita and Ryan Burnell and Bogdan Damoc and Junwhan Ahn and Andrew Brock and Mantas Pajarskas and Anastasia Petrushkina and Seb Noury and Lorenzo Blanco and Kevin Swersky and Arun Ahuja and Thi Avrahami and Vedant Misra and Raoul de Liedekerke and Mariko Iinuma and Alex Polozov and Sarah York and George van den Driessche and Paul Michel and Justin Chiu and Rory Blevins and Zach Gleicher and Adrià Recasens and Alban Rrustemi and Elena Gribovskaya and Aurko Roy and Wiktor Gworek and Sébastien M. R. Arnold and Lisa Lee and James Lee-Thorp and Marcello Maggioni and Enrique Piqueras and Kartikeya Badola and Sharad Vikram and Lucas Gonzalez and Anirudh Baddepudi and Evan Senter and Jacob Devlin and James Qin and Michael Azzam and Maja Trebacz and Martin Polacek and Kashyap Krishnakumar and Shuo-yiin Chang and Matthew Tung and Ivo Penchev and Rishabh Joshi and Kate Olszewska and Carrie Muir and Mateo Wirth and Ale Jakse Hartman and Josh Newlan and Sheleem Kashem and Vijay Bolina and Elahe Dabir and Joost van Amersfoort and Zafarali Ahmed and James Cobon-Kerr and Aishwarya Kamath and Arnar Mar Hrafnkelsson and Le Hou and Ian Mackinnon and Alexandre Frechette and Eric Noland and Xiance Si and Emanuel Taropa and Dong Li and Phil Crone and Anmol Gulati and Sébastien Cevey and Jonas Adler and Ada Ma and David Silver and Simon Tokumine and Richard Powell and Stephan Lee and Kiran Vodrahalli and Samer Hassan and Diana Mincu and Antoine Yang and Nir Levine and Jenny Brennan and Mingqiu Wang and Sarah Hodkinson and Jeffrey Zhao and Josh Lipschultz and Aedan Pope and Michael B. Chang and Cheng Li and Laurent El Shafey and Michela Paganini and Sholto Douglas and Bernd Bohnet and Fabio Pardo and Seth Odoom and Mihaela Rosca and Cicero Nogueira dos Santos and Kedar Soparkar and Arthur Guez and Tom Hudson and Steven Hansen and Chulayuth Asawaroengchai and Ravi Addanki and Tianhe Yu and Wojciech Stokowiec and Mina Khan and Justin Gilmer and Jaehoon Lee and Carrie Grimes Bostock and Keran Rong and Jonathan Caton and Pedram Pejman and Filip Pavetic and Geoff Brown and Vivek Sharma and Mario Lučić and Rajkumar Samuel and Josip Djolonga and Amol Mandhane and Lars Lowe Sjösund and Elena Buchatskaya and Elspeth White and Natalie Clay and Jiepu Jiang and Hyeontaek Lim and Ross Hemsley and Zeyncep Cankara and Jane Labanowski and Nicola De Cao and David Steiner and Sayed Hadi Hashemi and Jacob Austin and Anita Gergely and Tim Blyth and Joe Stanton and Kaushik Shivakumar and Aditya Siddhant and Anders Andreassen and Carlos Araya and Nikhil Sethi and Rakesh Shivanna and Steven Hand and Ankur Bapna and Ali Khodaei and Antoine Miech and Garrett Tanzer and Andy Swing and Shantanu Thakoor and Lora Aroyo and Zhufeng Pan and Zachary Nado and Jakub Sygnowski and Stephanie Winkler and Dian Yu and Mohammad Saleh and Loren Maggiore and Yamini Bansal and Xavier Garcia and Mehran Kazemi and Piyush Patil and Ishita Dasgupta and Iain Barr and Minh Giang and Thais Kagohara and Ivo Danihelka and Amit Marathe and Vladimir Feinberg and Mohamed Elhawaty and Nimesh Ghelani and Dan Horgan and Helen Miller and Lexi Walker and Richard Tanburn and Mukarram Tariq and Disha Shrivastava and Fei Xia and Qingze Wang and Chung-Cheng Chiu and Zoe Ashwood and Khuslen Baatarsukh and Sina Samangooei and Raphaël Lopez Kaufman and Fred Alcober and Axel Stjerngren and Paul Komarek and Katerina Tsihlas and Anudhyan Boral and Ramona Comanescu and Jeremy Chen and Ruibo Liu and Chris Welty and Dawn Bloxwich and Charlie Chen and Yanhua Sun and Fangxiaoyu Feng and Matthew Mauger and Xerxes Dotiwalla and Vincent Hellendoorn and Michael Sharman and Ivy Zheng and Krishna Haridasan and Gabe Barth-Maron and Craig Swanson and Dominika Rogozińska and Alek Andreev and Paul Kishan Rubenstein and Ruoxin Sang and Dan Hurt and Gamaleldin Elsayed and Renshen Wang and Dave Lacey and Anastasija Ilić and Yao Zhao and Adam Iwanicki and Alejandro Lince and Alexander Chen and Christina Lyu and Carl Lebsack and Jordan Griffith and Meenu Gaba and Paramjit Sandhu and Phil Chen and Anna Koop and Ravi Rajwar and Soheil Hassas Yeganeh and Solomon Chang and Rui Zhu and Soroush Radpour and Elnaz Davoodi and Ving Ian Lei and Yang Xu and Daniel Toyama and Constant Segal and Martin Wicke and Hanzhao Lin and Anna Bulanova and Adrià Puigdomènech Badia and Nemanja Rakićević and Pablo Sprechmann and Angelos Filos and Shaobo Hou and Víctor Campos and Nora Kassner and Devendra Sachan and Meire Fortunato and Chimezie Iwuanyanwu and Vitaly Nikolaev and Balaji Lakshminarayanan and Sadegh Jazayeri and Mani Varadarajan and Chetan Tekur and Doug Fritz and Misha Khalman and David Reitter and Kingshuk Dasgupta and Shourya Sarcar and Tina Ornduff and Javier Snaider and Fantine Huot and Johnson Jia and Rupert Kemp and Nejc Trdin and Anitha Vijayakumar and Lucy Kim and Christof Angermueller and Li Lao and Tianqi Liu and Haibin Zhang and David Engel and Somer Greene and Anaïs White and Jessica Austin and Lilly Taylor and Shereen Ashraf and Dangyi Liu and Maria Georgaki and Irene Cai and Yana Kulizhskaya and Sonam Goenka and Brennan Saeta and Ying Xu and Christian Frank and Dario de Cesare and Brona Robenek and Harry Richardson and Mahmoud Alnahlawi and Christopher Yew and Priya Ponnapalli and Marco Tagliasacchi and Alex Korchemniy and Yelin Kim and Dinghua Li and Bill Rosgen and Kyle Levin and Jeremy Wiesner and Praseem Banzal and Praveen Srinivasan and Hongkun Yu and Çağlar Ünlü and David Reid and Zora Tung and Daniel Finchelstein and Ravin Kumar and Andre Elisseeff and Jin Huang and Ming Zhang and Ricardo Aguilar and Mai Giménez and Jiawei Xia and Olivier Dousse and Willi Gierke and Damion Yates and Komal Jalan and Lu Li and Eri Latorre-Chimoto and Duc Dung Nguyen and Ken Durden and Praveen Kallakuri and Yaxin Liu and Matthew Johnson and Tomy Tsai and Alice Talbert and Jasmine Liu and Alexander Neitz and Chen Elkind and Marco Selvi and Mimi Jasarevic and Livio Baldini Soares and Albert Cui and Pidong Wang and Alek Wenjiao Wang and Xinyu Ye and Krystal Kallarackal and Lucia Loher and Hoi Lam and Josef Broder and Dan Holtmann-Rice and Nina Martin and Bramandia Ramadhana and Mrinal Shukla and Sujoy Basu and Abhi Mohan and Nick Fernando and Noah Fiedel and Kim Paterson and Hui Li and Ankush Garg and Jane Park and DongHyun Choi and Diane Wu and Sankalp Singh and Zhishuai Zhang and Amir Globerson and Lily Yu and John Carpenter and Félix de Chaumont Quitry and Carey Radebaugh and Chu-Cheng Lin and Alex Tudor and Prakash Shroff and Drew Garmon and Dayou Du and Neera Vats and Han Lu and Shariq Iqbal and Alex Yakubovich and Nilesh Tripuraneni and James Manyika and Haroon Qureshi and Nan Hua and Christel Ngani and Maria Abi Raad and Hannah Forbes and Jeff Stanway and Mukund Sundararajan and Victor Ungureanu and Colton Bishop and Yunjie Li and Balaji Venkatraman and Bo Li and Chloe Thornton and Salvatore Scellato and Nishesh Gupta and Yicheng Wang and Ian Tenney and Xihui Wu and Ashish Shenoy and Gabriel Carvajal and Diana Gage Wright and Ben Bariach and Zhuyun Xiao and Peter Hawkins and Sid Dalmia and Clement Farabet and Pedro Valenzuela and Quan Yuan and Ananth Agarwal and Mia Chen and Wooyeol Kim and Brice Hulse and Nandita Dukkipati and Adam Paszke and Andrew Bolt and Kiam Choo and Jennifer Beattie and Jennifer Prendki and Harsha Vashisht and Rebeca Santamaria-Fernandez and Luis C. Cobo and Jarek Wilkiewicz and David Madras and Ali Elqursh and Grant Uy and Kevin Ramirez and Matt Harvey and Tyler Liechty and Heiga Zen and Jeff Seibert and Clara Huiyi Hu and Andrey Khorlin and Maigo Le and Asaf Aharoni and Megan Li and Lily Wang and Sandeep Kumar and Norman Casagrande and Jay Hoover and Dalia El Badawy and David Soergel and Denis Vnukov and Matt Miecnikowski and Jiri Simsa and Praveen Kumar and Thibault Sellam and Daniel Vlasic and Samira Daruki and Nir Shabat and John Zhang and Guolong Su and Jiageng Zhang and Jeremiah Liu and Yi Sun and Evan Palmer and Alireza Ghaffarkhah and Xi Xiong and Victor Cotruta and Michael Fink and Lucas Dixon and Ashwin Sreevatsa and Adrian Goedeckemeyer and Alek Dimitriev and Mohsen Jafari and Remi Crocker and Nicholas FitzGerald and Aviral Kumar and Sanjay Ghemawat and Ivan Philips and Frederick Liu and Yannie Liang and Rachel Sterneck and Alena Repina and Marcus Wu and Laura Knight and Marin Georgiev and Hyo Lee and Harry Askham and Abhishek Chakladar and Annie Louis and Carl Crous and Hardie Cate and Dessie Petrova and Michael Quinn and Denese Owusu-Afriyie and Achintya Singhal and Nan Wei and Solomon Kim and Damien Vincent and Milad Nasr and Christopher A. Choquette-Choo and Reiko Tojo and Shawn Lu and Diego de Las Casas and Yuchung Cheng and Tolga Bolukbasi and Katherine Lee and Saaber Fatehi and Rajagopal Ananthanarayanan and Miteyan Patel and Charbel Kaed and Jing Li and Shreyas Rammohan Belle and Zhe Chen and Jaclyn Konzelmann and Siim Põder and Roopal Garg and Vinod Koverkathu and Adam Brown and Chris Dyer and Rosanne Liu and Azade Nova and Jun Xu and Alanna Walton and Alicia Parrish and Mark Epstein and Sara McCarthy and Slav Petrov and Demis Hassabis and Koray Kavukcuoglu and Jeffrey Dean and Oriol Vinyals},
      year={2024},
      eprint={2403.05530},
      archivePrefix={arXiv},
      primaryClass={cs.CL}
}

@misc{wei2023chainofthought,
      title={Chain-of-Thought Prompting Elicits Reasoning in Large Language Models}, 
      author={Jason Wei and Xuezhi Wang and Dale Schuurmans and Maarten Bosma and Brian Ichter and Fei Xia and Ed Chi and Quoc Le and Denny Zhou},
      year={2023},
      eprint={2201.11903},
      archivePrefix={arXiv},
      primaryClass={cs.CL}
}

@misc{suzgun2024metaprompting,
      title={Meta-Prompting: Enhancing Language Models with Task-Agnostic Scaffolding}, 
      author={Mirac Suzgun and Adam Tauman Kalai},
      year={2024},
      eprint={2401.12954},
      archivePrefix={arXiv},
      primaryClass={cs.CL}
}

@misc{wei2024jailbreak,
      title={Jailbreak and Guard Aligned Language Models with Only Few In-Context Demonstrations}, 
      author={Zeming Wei and Yifei Wang and Ang Li and Yichuan Mo and Yisen Wang},
      year={2024},
      eprint={2310.06387},
      archivePrefix={arXiv},
      primaryClass={cs.LG}
}

@misc{han2024parameterefficient,
      title={Parameter-Efficient Fine-Tuning for Large Models: A Comprehensive Survey}, 
      author={Zeyu Han and Chao Gao and Jinyang Liu and Jeff Zhang and Sai Qian Zhang},
      year={2024},
      eprint={2403.14608},
      archivePrefix={arXiv},
      primaryClass={cs.LG}
}

@misc{qi2023finetuning,
      title={Fine-tuning Aligned Language Models Compromises Safety, Even When Users Do Not Intend To!}, 
      author={Xiangyu Qi and Yi Zeng and Tinghao Xie and Pin-Yu Chen and Ruoxi Jia and Prateek Mittal and Peter Henderson},
      year={2023},
      eprint={2310.03693},
      archivePrefix={arXiv},
      primaryClass={cs.CL}
}

@article{fard2021distributing,
  title={Distributing and Democratizing Institutional Power Through Decentralization},
  author={Fard Bahreini, Amir and Collomosse, John and Seidel, Marc-David L and Sotoudehnia, Maral and Woo, Carson C},
  journal={Building Decentralized Trust: Multidisciplinary Perspectives on the Design of Blockchains and Distributed Ledgers},
  pages={95--109},
  year={2021},
  publisher={Springer}
}

@techreport{glaze2024,
    author = {Parsee Mizuhashi},
    title = {Glaze and the Effectiveness of Anti-AI Methods for Diffusion Models
(https://huggingface.co/blog/parsee-mizuhashi/glaze-and-anti-ai-methods)},
    institution = {HuggingFace},
    year = {May 15 2024}
}

@techreport{glitch,
    author = {Rumbelow, Jessica and Watkins, Matt},
    title = {SolidGoldMagikarp (plus, prompt generation)},
    institution = {LessWrong},
    year = {February 5 2023}
}

@article{ganguli2022red,
  title={Red teaming language models to reduce harms: Methods, scaling behaviors, and lessons learned},
  author={Ganguli, Deep and Lovitt, Liane and Kernion, Jackson and Askell, Amanda and Bai, Yuntao and Kadavath, Saurav and Mann, Ben and Perez, Ethan and Schiefer, Nicholas and Ndousse, Kamal and others},
  journal={arXiv preprint arXiv:2209.07858},
  year={2022}
}

@article{brynjolfsson2023big,
  title={Big AI can centralize decision-making and power, and that’sa problem},
  author={Brynjolfsson, Erik and Ng, Andrew},
  journal={Missing links in ai governance},
  volume={65},
  year={2023}
}

@article{gupta2020decentralization,
  title={Decentralization of artificial intelligence: Analyzing developments in decentralized learning and distributed AI networks},
  author={Gupta, Ishan},
  journal={arXiv preprint arXiv:1603.04467},
  year={2020}
}

@misc{akiba2024evolutionary,
      title={Evolutionary Optimization of Model Merging Recipes}, 
      author={Takuya Akiba and Makoto Shing and Yujin Tang and Qi Sun and David Ha},
      year={2024},
      eprint={2403.13187},
      archivePrefix={arXiv},
      primaryClass={cs.NE}
}

@techreport{deepmind2024fsf,
    author = {Deepmind},
    title = {Frontier Safety Framework},
    institution = {Google Deepmind},
    year = {17 May 2024}
}

@misc{christiano2023deep,
      title={Deep reinforcement learning from human preferences}, 
      author={Paul Christiano and Jan Leike and Tom B. Brown and Miljan Martic and Shane Legg and Dario Amodei},
      year={2023},
      eprint={1706.03741},
      archivePrefix={arXiv},
      primaryClass={stat.ML}
}

@techreport{clark2023375,
    author = {Clark , Jack},
    title = {Import AI Import AI 375: GPT-2 five years later; decentralized training; new ways of thinking about consciousness and AI
},
    institution = {Demoscene AI},
    year = {2024}
}

@techreport{akash2023,
    author = {Murty, Anil},
    title = {Distributed Machine Learning on Akash Network With Ray (https://akash.network/)},
    institution = {Akash},
    year = {January 28 2024}
}

@techreport{gensyn2024,
    author = {gensyn},
    title = {Gensyn Litepaper: The hyperscale, cost-efficient compute protocol for the world’s deep learning models(https://docs.gensyn.ai/litepaper)},
    institution = {Gensyn},
    year = {2024}
}

@misc{zhao2023lingualinked,
      title={LinguaLinked: A Distributed Large Language Model Inference System for Mobile Devices}, 
      author={Junchen Zhao and Yurun Song and Simeng Liu and Ian G. Harris and Sangeetha Abdu Jyothi},
      year={2023},
      eprint={2312.00388},
      archivePrefix={arXiv},
      primaryClass={cs.LG}
}

@techreport{garg2024,
    author = {Garg, Aksh},
    title = {Shard: On the decentralized training of foundation models (https://aksh-garg.medium.com/shard-on-the-decentralized-training-of-foundation-models-2fd982176724)},
    institution = {Medium},
    year = {May 20 2024}
}

@misc{chiang2024chatbot,
      title={Chatbot Arena: An Open Platform for Evaluating LLMs by Human Preference}, 
      author={Wei-Lin Chiang and Lianmin Zheng and Ying Sheng and Anastasios Nikolas Angelopoulos and Tianle Li and Dacheng Li and Hao Zhang and Banghua Zhu and Michael Jordan and Joseph E. Gonzalez and Ion Stoica},
      year={2024},
      eprint={2403.04132},
      archivePrefix={arXiv},
      primaryClass={cs.AI}
}

@techreport{dwarkesh2024zuck,
    author = {Patel, Dwarkesh},
    title = {Mark Zuckerberg - Llama 3, Open Sourcing 10b Models, Caesar Augustus},
    institution = {Dwarkesh podcast},
    year = {18 April 2024}
}

@techreport{cohere2024,
    author = {Cohere},
    title = {Command R+ (https://docs.cohere.com/docs/command-r-plus)},
    institution = {Cohere},
    year = {23 May 2024}
}

@techreport{github2021,
    author = {Github},
    title = {Github Copilot (https://github.com/features/copilot)},
    institution = {Github},
    year = {october 2021}
}

@inproceedings{wu2023autogen,
      title={AutoGen: Enabling Next-Gen LLM Applications via Multi-Agent Conversation Framework},
      author={Qingyun Wu and Gagan Bansal and Jieyu Zhang and Yiran Wu and Beibin Li and Erkang Zhu and Li Jiang and Xiaoyun Zhang and Shaokun Zhang and Jiale Liu and Ahmed Hassan Awadallah and Ryen W White and Doug Burger and Chi Wang},
      year={2023},
      eprint={2308.08155},
      archivePrefix={arXiv},
      primaryClass={cs.AI}
}

@article{zhang2024training,
  title={Training Language Model Agents without Modifying Language Models},
  author={Zhang, Shaokun and Zhang, Jieyu and Liu, Jiale and Song, Linxin and Wang, Chi and Krishna, Ranjay and Wu, Qingyun},
  journal={ICML'24},
  year={2024}
}

@techreport{meta2024,
    author = {Meta AI Research},
    title = {Introducing Meta Llama 3: The most capable openly available LLM to date (https://ai.meta.com/blog/meta-llama-3/)},
    institution = {Meta},
    year = {18 April 2024}
}

@techreport{harris2023howto,
    author = {Harris, David Evan},
    title = {How to Regulate Unsecured “Open-Source” AI: No Exemptions},
    institution = {Tech Policy Press},
    year = {18 December 2023}
}

@techreport{huggingface2023content,
    author = {HuggingFace},
    title = {Content Policy (https://huggingface.co/content-guidelines)},
    institution = {HuggingFace} ,
    year = {30 August 2023}
}

@article{montes2019distributed,
  title={Distributed, decentralized, and democratized artificial intelligence},
  author={Montes, Gabriel Axel and Goertzel, Ben},
  journal={Technological Forecasting and Social Change},
  volume={141},
  pages={354--358},
  year={2019},
  publisher={Elsevier}
}

@misc{kersic2024review,
      title={A Review on Building Blocks of Decentralized Artificial Intelligence}, 
      author={Vid Kersic and Muhamed Turkanovic},
      year={2024},
      eprint={2402.02885},
      archivePrefix={arXiv},
      primaryClass={cs.AI}
}

@article{lin2024decentralized,
  title={Decentralized Physical Infrastructure Network (DePIN): Challenges and Opportunities},
  author={Lin, Zhibin and Wang, Taotao and Shi, Long and Zhang, Shengli and Cao, Bin},
  journal={arXiv preprint arXiv:2406.02239},
  year={2024}
}

@article{gabriel2022toward,
  title={Toward a theory of justice for artificial intelligence},
  author={Gabriel, Iason},
  journal={Daedalus},
  volume={151},
  number={2},
  pages={218--231},
  year={2022},
  publisher={MIT Press One Rogers Street, Cambridge, MA 02142-1209, USA journals-info~…}
}

@article{kreps2023ai,
  title={How AI Threatens Democracy},
  author={Kreps, Sarah and Kriner, Doug},
  journal={Journal of Democracy},
  volume={34},
  number={4},
  pages={122--131},
  year={2023},
  publisher={Johns Hopkins University Press}
}

@article{manheim2019artificial,
  title={Artificial intelligence: Risks to privacy and democracy},
  author={Manheim, Karl and Kaplan, Lyric},
  journal={Yale JL and Tech.},
  volume={21},
  pages={106},
  year={2019},
  publisher={HeinOnline}
}

@article{bontridder2021role,
  title={The role of artificial intelligence in disinformation},
  author={Bontridder, No{\'e}mi and Poullet, Yves},
  journal={Data and Policy},
  volume={3},
  pages={e32},
  year={2021},
  publisher={Cambridge University Press}
}

@article{owid-internet,
    author = {Hannah Ritchie and Edouard Mathieu and Max Roser and Esteban Ortiz-Ospina},
    title = {Internet},
    journal = {Our World in Data},
    year = {2023},
   }

@article{statista2022,
    author = {Evans, David},
    title =  {Number of software developers worldwide in 2018 to 2024 (in millions)},
    journal = {Statista},
    year = {(August 21, 2023)}
}

@techreport{maslej2023report,
    author = {Maslej, N, et al.},
    title = {Artificial Intelligence Index Report 2023: Chapter 7 Diversity},
    institution = {Institute for HumanCentered AI, Stanford University, Stanford, CA},
    year = {2023}
}

@article{shevlane2022structured,
  title={Structured access: an emerging paradigm for safe AI deployment},
  author={Shevlane, Toby},
  journal={arXiv preprint arXiv:2201.05159},
  year={2022}
}

@inproceedings{solaiman2023gradient,
  title={The gradient of generative AI release: Methods and considerations},
  author={Solaiman, Irene},
  booktitle={Proceedings of the 2023 ACM conference on fairness, accountability, and transparency},
  pages={111--122},
  year={2023}
}

@article{bucknall2023structured,
  title={STRUCTURED ACCESS FOR THIRD-PARTY RESEARCH ON FRONTIER AI MODELS: INVESTIGATING RESEARCHERS’MODEL ACCESS REQUIREMENTS},
  author={Bucknall, Benjamin S and Trager, Robert F},
  year={2023}
}

@incollection{garfinkel2021does,
  title={How does the offense-defense balance scale?},
  author={Garfinkel, Ben and Dafoe, Allan},
  booktitle={Emerging Technologies and International Stability},
  pages={247--274},
  year={2021},
  publisher={Routledge}
}

@article{bernardi2024societal,
  title={Societal Adaptation to Advanced AI},
  author={Bernardi, Jamie and Mukobi, Gabriel and Greaves, Hilary and Heim, Lennart and Anderljung, Markus},
  journal={arXiv preprint arXiv:2405.10295},
  year={2024}
}

@article{nissenbaum2004privacy,
author = {Nissenbaum, Helen},
year = {2004},
month = {05},
pages = {},
title = {Privacy As Contextual Integrity},
volume = {79},
journal = {Washington Law Review}
}

@techreport{deepbrainchain,
    author = {DeepBrain Chain},
    title = {Welcome to DeepBrain Chain (https://www.deepbrainchain.org/)},
    institution = {DeepBrain Chain},
    year = {2017}
}

@inproceedings{10.1145/3339252.3340516,
author = {Hor\'{a}k, Martin and Stupka, V\'{a}clav and Hus\'{a}k, Martin},
title = {GDPR Compliance in Cybersecurity Software: A Case Study of DPIA in Information Sharing Platform},
year = {2019},
isbn = {9781450371643},
publisher = {Association for Computing Machinery},
address = {New York, NY, USA},
url = {https://doi.org/10.1145/3339252.3340516},
doi = {10.1145/3339252.3340516},
booktitle = {Proceedings of the 14th International Conference on Availability, Reliability and Security},
articleno = {36},
numpages = {8},
location = {Canterbury, CA, United Kingdom},
series = {ARES '19}
}

@article{bruun2018artificial,
  title={Artificial intelligence, jobs and the future of work: Racing with the machines},
  author={Bruun, Edvard PG and Duka, Alban},
  journal={Basic Income Studies},
  volume={13},
  number={2},
  pages={20180018},
  year={2018},
  publisher={De Gruyter}
}

@techreport{huggingfaceindex,
    author = {HuggingFace},
    title = {HugginFace index (https://huggingface.co/docs/hub/en/index)},
    institution = {Huggingface},
    year = {2024} 
}

@techreport{iexec,
    author = {iExec},
    title = {Build, Own and Monetise Web3 (https://iex.ec/)},
    institution = {iExec},
    year = {2020} 
}

@techreport{golem,
    author = {Golem},
    title = {Golem (https://www.golem.network/)} ,
    institution = {Golem},
    year = {2021}
}

@inproceedings{agarwal2024multi,
  title={Multi-Stage Prompting for Next Best Agent Recommendations in Adaptive Workflows},
  author={Agarwal, Prerna and Dave, Harshit and Bandlamudi, Jayachandu and Sindhgatta, Renuka and Mukherjee, Kushal},
  booktitle={Proceedings of the AAAI Conference on Artificial Intelligence},
  volume={38},
  number={21},
  pages={22843--22849},
  year={2024}
}

@techreport{dwarkeshtrenton,
    author = {Patel, Dwarkesh},
    title = {Sholto Douglas and Trenton Bricken - How to Build and Understand GPT-7's Mind (https://www.dwarkeshpatel.com/p/sholto-douglas-trenton-bricken)},
    institution = {Dwarkesh Podcast},
    year = {28 May 2024}
}

@techreport{rsp,
    author = {METR},
    title = {Responsible Scaling Policies (https://metr.org/blog/2023-09-26-rsp/)},
    institution = {METR},
    year = {26 September 2023 
}
}

@techreport{openaiprepared,
    author = {OpenAI},
    title = {Preparedness Framework (Beta)},
    institution = {OpenAI},
    year = {December 18 2023}
}

@techreport{googlesafety,
    author = {Dragan, Anca and King, Helen and Dafoe, Allan},
    title = {Introducing the Frontier Safety Framework},
    institution = {Google DeepMind},
    year = {2024}
}

@misc{egan2023oversight,
      title={Oversight for Frontier AI through a Know-Your-Customer Scheme for Compute Providers}, 
      author={Janet Egan and Lennart Heim},
      year={2023},
      eprint={2310.13625},
      archivePrefix={arXiv},
      primaryClass={cs.CY}
}

@article{gordon2022mapping,
  title={On mapping values in AI governance},
  author={Gordon, Geoff and Rieder, Bernhard and Sileno, Giovanni},
  journal={Computer, Law and Security Review},
  volume={46},
  pages={105712},
  year={2022},
  publisher={Elsevier}
}

@article{mercille2019market,
  title={Market, non-market and anti-market processes in neoliberalism},
  author={Mercille, Julien and Murphy, Enda},
  journal={Critical Sociology},
  volume={45},
  number={7-8},
  pages={1093--1109},
  year={2019},
  publisher={SAGE Publications Sage UK: London, England}
}

@article{donnelly2023semiconductor,
  title={Semiconductor and ICT Industrial Policy in the US and EU: Geopolitical Threat Responses},
  author={Donnelly, Shawn},
  journal={Politics and Governance},
  volume={11},
  number={4},
  pages={129--139},
  year={2023}
}

@article{aarne2024secure,
    author = {Aarne, Onni and Fist, Tim and Withers, Caleb},
    title = {Secure, Governable Chips Using On-Chip Mechanisms to Manage National Security Risks from AI and Advanced Computing},
    journal = {CNAS},
    year = {January 2024}
}

@article{prabhakaran2022human,
  title={A human rights-based approach to responsible AI},
  author={Prabhakaran, Vinodkumar and Mitchell, Margaret and Gebru, Timnit and Gabriel, Iason},
  journal={arXiv preprint arXiv:2210.02667},
  year={2022}
}

@article{robertson1977scientist,
  title={The Scientist's Rights to Research: A Constitutional Analysis},
  author={Robertson, John A},
  journal={S. Cal. l. Rev.},
  volume={51},
  pages={1203},
  year={1977},
  publisher={HeinOnline}
}

@article{verschraegen2018regulating,
  title={Regulating scientific research: A constitutional moment?},
  author={Verschraegen, Gert},
  journal={Journal of Law and Society},
  volume={45},
  pages={S163--S184},
  year={2018},
  publisher={Wiley Online Library}
}

@article{goldberg1979constitutional,
  title={The Constitutional Status of American Science},
  author={Goldberg, Steven},
  journal={U. Ill. LF},
  pages={1},
  year={1979},
  publisher={HeinOnline}
}

@article{nevo2024securing,
  title={RAND Report: Securing AI Model Weights},
  author={Nevo, Sella and Lahav, Dan and Karpur, Ajay and Bar-on, Yogev and Bradley, Henry Alexander and Alstott, Jeff},
  year={2024}
}

@techreport{NIST,
    author = {Computer Security Division, Information Technology Laboratory},
    title = {Managing Information Security Risk Organization, Mission, and Information System View },
    institution = {National Institute of Standards and Technology},
    year = {2017}
}

@article{lewis2019information,
  title={Information hazards in biotechnology},
  author={Lewis, Gregory and Millett, Piers and Sandberg, Anders and Snyder-Beattie, Andrew and Gronvall, Gigi},
  journal={Risk Analysis},
  volume={39},
  number={5},
  pages={975--981},
  year={2019},
  publisher={Wiley Online Library}
}

@article{atlas2005,
    author = {Atlas, RM},
    title = {Biodefense research: an emerging conundrum},
    journal = {Current Opinion Biotechnology},
    year = {2005 June}
}

@article{noyce2018construction,
  title={Construction of an infectious horsepox virus vaccine from chemically synthesized DNA fragments},
  author={Noyce, Ryan S and Lederman, Seth and Evans, David H},
  journal={PloS one},
  volume={13},
  number={1},
  pages={e0188453},
  year={2018},
  publisher={Public Library of Science San Francisco, CA USA}
}

@article{imai2012experimental,
  title={Experimental adaptation of an influenza H5 HA confers respiratory droplet transmission to a reassortant H5 HA/H1N1 virus in ferrets},
  author={Imai, Masaki and Watanabe, Tokiko and Hatta, Masato and Das, Subash C and Ozawa, Makoto and Shinya, Kyoko and Zhong, Gongxun and Hanson, Anthony and Katsura, Hiroaki and Watanabe, Shinji and others},
  journal={Nature},
  volume={486},
  number={7403},
  pages={420--428},
  year={2012},
  publisher={Nature Publishing Group UK London}
}

@techreport{conjecture,
    author = {Leahy, Connor and Black, Sid and Scammell, Chris and Miotti, Andrea},
    title = {Conjecture Internal Infohazard Policy},
    institution = {Conjecture},
    year = {29th July 2022}
}

@misc{goldstein2023understanding,
      title={Understanding the (In)Effectiveness of Content Moderation: A Case Study of Facebook in the Context of the U.S. Capitol Riot}, 
      author={Ian Goldstein and Laura Edelson and Minh-Kha Nguyen and Oana Goga and Damon McCoy and Tobias Lauinger},
      year={2023},
      eprint={2301.02737},
      archivePrefix={arXiv},
      primaryClass={cs.SI}
}

@article{nytimes,
    author = {Popli, Nik},
    title = {The 5 Most Important Revelations From the ‘Facebook Papers’
 https://time.com/6110234/facebook-papers-testimony-explained/},
    journal = {Time},
    year = {October 26, 2021}
}

@article{greaves2021case,
  title={The case for strong longtermism},
  author={Greaves, Hilary and MacAskill, William},
  journal={Global Priorities Institute Working Paper No. 5-2021},
  year={2021}
}

@article{roberts2024global,
  title={Global AI governance: barriers and pathways forward},
  author={Roberts, Huw and Hine, Emmie and Taddeo, Mariarosaria and Floridi, Luciano},
  journal={International Affairs},
  volume={100},
  number={3},
  pages={1275--1286},
  year={2024},
  publisher={Oxford University Press}
}

@article{daly2020ai,
  title={AI, governance and ethics: global perspectives},
  author={Daly, Angela and Hagendorff, Thilo and Li, Hui and Mann, Monique and Marda, Vidushi and Wagner, Ben and Wang, Wayne Wei},
  journal={University of Hong Kong Faculty of Law Research Paper},
  number={2020/051},
  year={2020}
}

@article{toufaily2021framework,
  title={A framework of blockchain technology adoption: An investigation of challenges and expected value},
  author={Toufaily, Elissar and Zalan, Tatiana and Dhaou, Soumaya Ben},
  journal={Information and Management},
  volume={58},
  number={3},
  pages={103444},
  year={2021},
  publisher={Elsevier}
}

@article{ciarli2021digital,
  title={Digital technologies, innovation, and skills: Emerging trajectories and challenges},
  author={Ciarli, Tommaso and Kenney, Martin and Massini, Silvia and Piscitello, Lucia},
  journal={Research Policy},
  volume={50},
  number={7},
  pages={104289},
  year={2021},
  publisher={Elsevier}
}

@misc{song2024position,
      title={Position: Leverage Foundational Models for Black-Box Optimization}, 
      author={Xingyou Song and Yingtao Tian and Robert Tjarko Lange and Chansoo Lee and Yujin Tang and Yutian Chen},
      year={2024},
      eprint={2405.03547},
      archivePrefix={arXiv},
}

@techreport{bittererlesson,
    author = {McLaughlin, Aidan},
    title = {The Bitter-er Lesson},
    institution = {Personal Blog},
    year = {14 June 2024}
}

@techreport{Situational,
    author = {Aschenbrenner, Leopold},
    title = {Racing to the Trillion-Dollar Cluster},
    institution = {Situational Awareness (Blog)},
    year = {June 6th 2024}
}

@incollection{martin2020ethical,
  title={Ethical issues in the big data industry},
  author={Martin, Kirsten E},
  booktitle={Strategic Information Management},
  pages={450--471},
  year={2020},
  publisher={Routledge}
}

@techreport{davidevan,
    author = {Evan Harris, David},
    title = {How to Regulate Unsecured “Open-Source” AI: No Exemptions},
    institution = {Tech Policy Press},
    year = {December 4 2023}
}

\end{document}